\documentclass
[preprint,pra,a4paper,showpacs,showkeys,byrevtex,nobibnotes,nofootinbib,10pt]{revtex4}%
\usepackage{amsfonts}
\usepackage{amsmath}
\usepackage{amssymb}
\usepackage{graphicx}%
\providecommand{\U}[1]{\protect\rule{.1in}{.1in}}

\begin{document}
\title{Interpretations of Quantum Theory in the Light of Modern Cosmology}
\author{Mario Castagnino$^{1}$}
\author{Sebastian Fortin$^{2}$}
\author{Roberto Laura$^{3}$}
\author{Daniel Sudarsky$^{4}$}

\begin{abstract}
The difficult issues related to the interpretation of quantum mechanics and,
in particular, the \textquotedblleft measurement problem" are revisited using
as motivation the process of generation of structure from quantum fluctuations
in inflationary cosmology. The unessential mathematical complexity of the
particular problem is bypassed, facilitating the discussion of the conceptual
issues, by considering, within the paradigm set up by the cosmological
problem, another problem where symmetry serves as a focal point: a simplified
version of Mott's problem.

\end{abstract}
\preprint{ICN-UNAM-05/01}
\affiliation{$^{1}$Instituto de Astronom\'{\i}a y F\'{\i}sica del Espacio (CONICET-UBA) and
Instituto de F\'{\i}sica Rosario (CONICET-UNR), Argentina.}
\affiliation{$^{2}$CONICET, Departamento de F\'{\i}sica FCEN (Universidad de Buenos Aires), Argentina.}
\affiliation{$^{3}$Instituto de F\'{\i}sica Rosario (CONICET-UNR) and Facultad de Ciencias
Exactas, Ingenier\'{\i}a y Agrimensura (UNR), Pellegrini 250, 2000 Rosario, Argentina.}
\affiliation{$^{4}$Instituto de Ciencias Nucleares, Universidad Nacional Aut\'{o}noma de
M\'{e}xico, A. Postal 70-543, M\'{e}xico D.F. 04510, M\'{e}xico.}
\maketitle
\tableofcontents

\section{Introduction}

It is a remarkable fact that the debate about the interpretation of quantum
mechanics continues more than 80 years after the establishment of that theory
in its modern form. This is due in part to the fact that the theory is
extremely successful, and that the multiple interpretations seem to lead to
exactly the same predictions\footnote{We are ignoring the fact that certain
interpretations are problematic.The point however is that to the extent that
they are applied in a particular manner in concrete situations they do not
offer predictions that differ from the text book version of Quantum Theory.}
when applied to all the situations we have faced until now. In other words,
when faced with any laboratory situation, one can rely on any of the
interpretations, as they all lead, in practice, to exactly the same answers
and predictions regarding the observations.

We will see that the situation changes dramatically when confronted with the
challenges posed by modern cosmology. We will argue that, in that case, none
of the existing interpretations are sufficient to deal successfully with the
problems at hand.

We must face serious problems even before one gets into full quantum
cosmology, where contact with observation is more elusive than in the case we
will be focussing on. In fact, once one tries to incorporate gravitation into
the quantum treatment, and quite independently of the technical issues that
must be confronted, this situation entails yet another set of very serious
conceptual problems, such as the disappearance of time from the theory
\cite{QGTime}, and many others \cite{Conceptual Problems of QG}.

The issue we want to consider here is one that arises already when considering
the inflationary regime that, according to the current understanding, is an
essential aspect of the history of our universe. We note that this situation
is one where the technical difficulties associated with a full quantum theory
of gravitation are essentially absent and simple perturbative treatments seems
to be sufficient. We will see however that despite the relative simplicity of
the situation, a serious question must be confronted.

Let us start by recalling here that the inflationary modification or
adjustment to our cosmological theories arose when attempting to deal with
certain \textquotedblleft naturalness problems" of the standard Hot Big Bang
Model: Namely, the Horizon problem, The Flatness problem and the Primordial
relics problem \cite{Problems Big Bang}. The inflationary solution is obtained
when one assumes that the \textquotedblleft standard cosmological
era\textquotedblright\ is preceded by an era of almost exponential expansion,
which erases all inhomogeneities, dissolves all defects, and, in general,
drives all quantum fields to their vacuum state. Although inflation was
introduced to deal with those naturalness problems in the standard Big Bang
Cosmological theory, its major success is its purported ability to predict the
shape of the spectrum of primordial fluctuations that are supposed to seed all
the structure in our universe, and whose earliest manifestations we see
imprinted in the Cosmic Microwave Background (CMB).

The problem we want to focus on, is exactly how does our theory account for
the manner in which those first seeds of structure actually emerge from the
quantum fluctuations of the inflaton field\footnote{The favored version of the
theory actually deals with a composite variable representing the quantum
aspects of the inflaton field and a certain component of the space-time
metric\cite{Muckhanov}.}. We will see that, although the problem is, in a
sense, connected with the measurement problem in Quantum Theory, the
particular manner in which it occurs in the inflationary context is such that
issues which otherwise one might consider as having "only philosophical
relevance", become acute to the point that a major shift in our thinking is required.

The core of the problem can be summarized in the following question:
\textquotedblleft How is it that from an initial situation which is supposed
to be described, both at the quantum and classical levels, by
conditions\footnote{We refer here to the stage corresponding to several
e-folds after the start of inflation, when the background corresponds to an
inflating, flat, Robertson Walker space-time, and the \textquotedblleft
quantum fluctuations" are described by the Bunch-Davies vacuum, or some
similarly highly symmetric state. This characterization is thought to be
accurate up to exponentially small corrections in the number of e-folds, a
detail that we will ignore as is customary in all inflationary analyses.} that
are perfectly homogeneous and isotropic, a universe with space-time dependent
density perturbations emerges, through processes that involve only dynamics
which does not break the initial symmetry?

The issue has been confronted by several researchers in the field of
Inflationary Cosmology, and it is worth mentioning that the majority of
colleagues working on that subject do not seem to think that there is a
problem, or are convinced that the problem has been solved by some clever
arguments. It is noteworthy however, that these arguments tend to differ, in
general, from one inflationary cosmologist to another \cite{Cosmologists}.
Other cosmologists do acknowledge that there seems to be something unclear at
this point \cite{Padmanabhan}, and the work of \cite{Trashen} might be
considered as an early inquiry on the subject. Moreover, a couple of recent
books on the subject acknowledge that there is a problem (see \cite{Weinberg}
and \cite{Muckhanov}).

The issue has been mostly ignored also by the community working in
Foundational issues in Quantum Theory. They are probably justified in thinking
that the complexity of the cosmological situation, involving as it does, not
only general relativity but also quantum field theory in curved space-time, is
not a particularly convenient one to consider in dealing with fundamental and
conceptual questions. We believe, however, that the issue we have just
described, actually offers an opportunity to focus sharply on the problems
that normally concern our colleagues in that field, and that important lessons
can be extracted by considering the issues in some detail. This manuscript is
devoted precisely towards that goal.

Our strategy here will be to find a simpler \textquotedblleft analogous"
situation where the relevant issues appear just as in the cosmological
setting, but where we have removed the complications that usually hide the
fundamental aspects we want to focus on.

The paper will be organized as follows: In section II, we will review the
essential aspects and details of the cosmological problem as it is treated in
the works on inflationary cosmology. In section III, we will discuss briefly a
problem that is often presented as analogous to the one we are confronting,
the problem of breaking of rotational symmetry in the observations of nuclear
decay in bubble chambers (often called Mott`s problem \cite{Mott}), a problem
that is usually considered as solved in that work by Sir N. F. Mott, and, we
will examine to which degree the analogy holds and fails to hold, and to what
degree the problem has been truly solved. In addition, we will present an even
simpler version of Mott's problem (which we call the Mini-Mott problem), that
will allow us to write all expressions in full detail, and thus, to focus more
clearly on the issue we must confront. In Section IV, we will then analyze the
manner in which the problem would have been addressed by a scientist adhering
to each one of the existing interpretational schemes for quantum theory. We
end with a brief discussion of our findings, and a conclusion.

\section{The Problem}

In presenting the basic aspects of the problem here, we will be ignoring
alternative views associated with certain interpretations of quantum theory,
that will be discussed in more detail in the rest of the manuscript. This is
done for clarity of exposition only, as is it not our aim to avoid the
discussion of relevant postures.

For most of its existence, cosmology has been discussed in a classical
language, as it is, in fact done in many other situations, such as the study
of trajectories of space probes, while everybody knows that our world is
quantum mechanical\footnote{There are, apparently, some people who disagree
with this view, but we will not consider their thinking any further here.}.
We, physicists, believe that the classical description of any system is
nothing but an approximation to the truly fundamental quantum description,
and, therefore, when we consider say, the classical description of the
trajectory of a satellite in space, we view it as indicating that the wave
function of its constituting atoms (or even that of its more elementary
constituents) is a sharply peaked wave packet, where the uncertainties in the
position and velocities are negligible compared with the precision of the
description we are making.\footnote{It even seems possible to construct wave
packets with high $n$ in an hydrogen atom that resemble to some degree the
situation above.} In those situations, the classical description does not
enter into a fundamental contradiction with the characteristics of our
satellite trajectory. However, it would be very unsettling, for instance, if
we were forced to consider at the same time, the classical elliptical
trajectory of the satellite around Earth, while on the other hand we were
forced to admit that, at the more fundamental level, the satellite was
described by a spherically symmetric wave function. We know this is not the
case\footnote{There are apparently philosophical views inspired in Kantian
ontology where this statement could be questioned.}, and that the precise
quantum description of the situation would indeed correspond to a suitable
superposition of energy and momentum eigenfunctions leading to a wave packet
corresponding to a sharply localized object. Of course, the precise way to do
this faces, at this time, technically insurmountable problems; however, the
principle is clear. In fact, we also must recognize that, in the case of the
satellite, one is dealing with an open system, and its interaction with a
clearly identifiable environment,-- and the ensuing decoherence-- , is likely
to play an important role in making compatible the quantum and classical
descriptions \cite{decoherence}. At this point we should note that, despite
the widespread beliefs to the contrary, decoherence can not be claimed to
truly solve the measurement problem\footnote{In fact in order to do that one
would need not only to define the privileged basis but also to add a postulate
about actualization.} \cite{NO-SOL- MEASUREMENT}).

In the cosmological setting, however, when we want to connect our classical
descriptions of the cosmological late times, with say a quantum description of
the early cosmological eras, we should seek, in a similar manner, to address
the corresponding issues. That is, when considering the classical description
we must regard it as nothing but the shorthand for the essential
characteristics ( i.e., the values corresponding to peaks of the wave
functions) of a full quantum mechanical description.

The universe that we inhabit today is certainly very well described at the
classical level by an in-homogeneous and anisotropic classical state, and thus
we must consider that such description, is, in accordance with the previous
paragraph, nothing but a concise and imperfect characterization of an equally
in-homogeneous and anisotropic quantum state, where the wave functions are
peaked at those values of the variables corresponding to those indicated by
the classical description. This would, in principle, involve no essential
differences from the case of the classical and quantum description of our
satellite, except for the lack of a clearly identifiable environment, given
that we take the universe to include, by definition, all the degrees of
freedom of our theory. However, there is nothing that indicates that, even
without the identification of an environment, we should not be able to make,
in principle, such quantum semi-classical description through the use of the
sharply peaked wave functions and taking into account all the interactions in
the analysis of its dynamics. The situation changes dramatically, however, if
we want to seriously consider a theory, in which the early quantum state of
the universe was particularly simple in a very special and precise way. This
is the case in the inflationary paradigm, and in particular as it refers to
the predictions about the spectrum of perturbations that, in that paradigm,
are believed to arise from the uncertainties or fluctuations characterizing
the quantum state of the inflaton, and which, according to these ideas,
constitute the seeds of cosmic structure of our universe today.

Let us remind the reader of the basic mechanism by which inflation is meant to
deal with the \textquotedblleft naturalness problems" of standard Big Bang
Cosmology discussed in the introduction.The essential idea is that if the
Universe undergoes an early epoch of accelerated (almost exponential)
expansion (lasting at least some 80 e-folds), it would come out of this period
as an essentially flat and homogeneous space-time with an extreme dilution of
all relics and, indeed, of all particle species. The states of all fields
would thus be extremely well described by suitable vacua. The deviations from
this state will be exponentially small (with the exponent characterized by the
number of e-folds). What is required to achieve this, is something that
behaves early on as a cosmological constant, but that is later
\textquotedblleft turned off" as a result of its own dynamics, returning the
universe to the standard Big Bang cosmological evolutionary path. This is
generically thought to be the result of a scalar field with a potential of
certain specific characteristics called the \textquotedblleft inflaton field".
The remarkable fact is that this scheme also results in the perdition of a
spectrum of primordial quantum uncertainties of the inflaton field that
matches the form of the famous Harrison-Z{}'eldovich\cite{HZ} spectrum of
primordial perturbations and which has been observed in the multiple analysis
of the extraordinary data on the CMB sky collected in the various recent
experiments\cite{CMB exp}.

This is the basis of the claim that inflation \textquotedblleft accounts for
the seeds of the cosmic structure". They \textquotedblleft emerge from the
quantum vacuum", continue to evolve after inflation has ended, and after
leaving their mark on the CMB, result in the emergence of the structure of our
universe. That structure which at late times is characterized by galaxy
clusters, galaxies, stars, planets and, later on, is tied to the development
of the conditions permitting our own existence.

The issue we must face is: Can any of the interpretations of quantum theory be
consistently used to justify the standard inflationary scenario, by which a
simple state that is supposed to characterize the early state of the universe
in terms of the vacuum state of all fields\footnote{Except, of course the zero
modo of the inflaton.}, and the flat FRW space-time, and which corresponds to
a situation that is completely homogeneous and isotropic, would lead to the
anisotropic and inhomogeneous universe in which we live\footnote{This point is
sometimes characterized as the \textquotedblleft transition from the quantum
regime to the classical regime", but we find this a bit misleading: most
people would agree that there are no classical or quantum regimes. The
fundamental description ought to be always a quantum description. However,
there exist regimes in which certain quantities can be described to a
sufficient accuracy by their classical counterparts represented by the
corresponding expectation values. All this depends, of course, on the physical
state, the underlying dynamics, the quantity of interest, and the context
\textbf{which one is considering }}.

This article will be devoted, to a large extent, to deal with the conceptual
issues above, and will not include the developments that are possible when
adding new elements to deal with the shortcomings we encountered in the
present context. We refer the readers interested on those matters to previous
works \cite{Us, UsPhen}.

Here, we want to consider the most popular interpretations of quantum theory,
and analyze their usefulness in dealing with those problems in cosmology. On
the other hand, we will be focussing for the most part of the article, and for
simplicity of the discussion on a much simpler example, where all calculations
can be made explicitly, however,in order to ensure that the lessons from one
case can be used in the other, we will be forcing ourselves to avoid, in the
corresponding treatment, the use of any element that would be absent in the
cosmological situation which motivates our study.

\section{Mott's problem}

How does a quantum system lose a symmetry present in the initial state if the
interactions do not break it? Historically,it seems, the first time that this
issue was faced within the newly formulated quantum theory concerned the decay
of an excited atom or nucleus, from a spherical symmetric state, to an
unexcited nucleus or atom and an emitted particle usually taken---and in fact
observed --- to be escaping along a particular direction, which is clearly not
a spherically symmetric state of affairs. The issue is whether or not this can
be fully accounted for within quantum theory.

The problem was considered in \cite{Mott} in early days of quantum theory, and
its treatment is thought, by many colleagues, to have clarified the issue
completely. However, let us look at it a new: The setting considered consists
of a nucleus located at the origin of spatial cartesian coordinates ($\vec
{X}=\vec{0}$) in an excited (unstable) state $\left\vert \Psi^{+}\right\rangle
$ which is spherically symmetric, and ready to decay into an unexcited nucleus
$\left\vert \Psi^{0}\right\rangle $, plus an $\alpha$ particle in state
$\left\vert \Xi_{\alpha}\right\rangle $, which is also spherically symmetric.
The setting includes also two hydrogen atoms with their nuclei fixed at
positions $\vec{a}_{1}$and $\vec{a}_{2}$, and their corresponding electrons,
in the corresponding ground states. The issue that is discussed is the degree
to which the nuclei should be aligned with the origin ( i.e $\vec{a}_{2}%
=c\vec{a}_{1}$with $c$ real) if both atoms are to be excited by the outgoing
$\alpha$ particle.

The analysis indicates that the probability of both atoms getting excited is
significant only when there is a large degree of alignment, thus explaining
the fact that the $\alpha$ particle traces straight paths in a bubble chamber.

Thus, one might think that one has an example in which an initial state
possessing spherical symmetry $\left\vert \Psi^{+}\right\rangle $ evolves into
a final state lacking such symmetry, despite the assumption that the
hamiltonian (governing the decay $\left\vert \Psi^{+}\right\rangle
\rightarrow\left\vert \Psi^{0}\right\rangle \left\vert \Xi_{\alpha
}\right\rangle $ and the $\alpha$ particle evolution) is symmetric under
rotations. Thus the problem would seemed to have disappeared and the
contradictory conclusions seemed to have vanished without trace. This seems
quite remarkable indeed.

However, a closer look reveals the fallacy: As indicated, the setting includes
the two unexcited atoms, which, through the localizations of their nuclei,
break the rotational symmetry. Indeed, the discussion is based, not on the
Hamiltonian for the evolution of the free $\alpha$ particle, but rather on the
Hamiltonian for the joint evolution (including the interaction) of the
$\alpha$ particle and the two electrons corresponding to the two localized
hydrogen atoms. In fact, the projection postulate associated with a
measurement is also coming into play in the analysis of \cite{Mott} when
computing probabilities by projecting on the subspace corresponding to the two
atoms being excited. It is clear that if we were to replace these atoms by
some hypothetical detectors whose quantum description corresponded to
spherically symmetric wave functions , each one with support, say, on a thin
spherical shell with radius $r_{i}$, a similar calculation would not lead to
straight lines, but rather it would lead us to expect a spherical pattern of
excitations. We would simply find that there was a certain probability for the
detectors corresponding to the shells $i^{th}\&j^{th}$ being excited, and the
symmetry would not have been compromised.

\subsection{An even simpler problem: Mini-Mott}

In order to deal with the mainly conceptual issues that confront us here, we
can make use of an even simpler problem where the symmetry in question is the
discrete spatial inversion in 1+1 dimensions. The problem consists of a free
non-relativistic particle of mass $M$ moving on a line and interacting with
suitable detectors located at two fixed points.

Consider a particle and two detectors with levels $|-\rangle$ (un-excited)
$|+\rangle$ ( excited ) located at $x=x_{0}$ and $x=-x_{0}$. Initially the
detectors are unexcited and the particle's wave function $\varphi
(x,0)=\left\langle x|\varphi_{0}\right\rangle $ is a simple gaussian centered
at $x=0$.

The Hamiltonian is:
\[
\hat{H}_{P}=\frac{1}{2M}{\hat{p}}^{2}%
\]
for the free particle part. The free hamiltonian for the detector located at
$x=+x_{0}$ is%

\begin{equation}
\hat{H}_{1}=\varepsilon|+\rangle{}_{1}\langle+|{}_{1}-\varepsilon|-\rangle
{}_{1}\langle-|{}_{1}%
\end{equation}
being $+\varepsilon$ ($-\varepsilon$) the energy of the detector in the exited
(unexcited) state $|+\rangle{}_{1}$ ($| - \rangle{}_{1}$).

The free Hamiltonian for the detector located at $x=-x_{0}$ is%

\begin{equation}
\hat{H}_{2}=\varepsilon|+\rangle{}_{2}\langle+|{}_{2}-\varepsilon|-\rangle
{}_{2}\langle-|{}_{2}%
\end{equation}

The Hamiltonian corresponding to the interaction of the particle and the
detector located at $x=+x_{0}$ is%

\begin{equation}
\hat{H}_{1P}=\lambda g(\hat{x}-x_{0}\hat{I_{p}})\otimes(|+\rangle{}_{1}%
\langle-|{}_{1}+|-\rangle{}_{1}\langle+|{}_{1})
\end{equation}
where $\hat{x}$ is the position operator of the particle, $\hat{I}_{p}$ the
identity in the Hilbert space of the particle $\left(  \hat{I}_{p}=\int
dx\left\vert x\right\rangle \left\langle x\right\vert \right)  $, and $g(y)$
is a function with support in a small interval centered at $y=0$
\cite{ScullySheaCullen}. Analogously, the Hamiltonian for the interaction of
the particle and the detector located at $x=-x_{0}$ is%
\[
\hat{H}_{2P}=\lambda g(\hat{x}+x_{0}\hat{I_{p}})\otimes(|+\rangle{}_{2}%
\langle-|{}_{2}+|-\rangle{}_{2}\langle+|{}_{2})
\]
The total Hamiltonian for the system composed by the particle and the two
detectors is%
\begin{align}
\hat{H}  &  =\hat{H}_{P}\otimes\hat{I}_{1}\otimes\hat{I}_{2}+\hat{I}%
_{P}\otimes\hat{H}_{1}\otimes\hat{I}_{2}+\hat{I}_{P}\otimes\hat{I}_{1}%
\otimes\hat{H}_{2}+\nonumber\\
&  +\hat{H}_{1P}\otimes\hat{I}_{2}+\hat{H}_{2P}\otimes\hat{I}_{1}
\label{ROB-01}%
\end{align}
where $\hat{I}_{1}\equiv|+\rangle{}_{1}\langle+|{}_{1}+|-\rangle{}_{1}%
\langle-|{}_{1}$ and $\hat{I}_{2}\equiv|+\rangle{}_{2}\langle+|{}%
_{2}+|-\rangle{}_{2}\langle-|{}_{2}$.

The Schr\"odinger equation can be solved explicitly with the initial condition%

\begin{equation}
\left\vert \Psi(t=0)\right\rangle =|\varphi_{0}\rangle\otimes|-\rangle{}%
_{1}\otimes|-\rangle{}_{2} \label{ROB-02}%
\end{equation}

Recall that the initial wave function for the particle is symmetric, so
$<-x|\varphi_{0}>=<\varphi_{0}|x>$. Thus, at time $t$ we have:%
\begin{align}
\left\vert \Psi(t)\right\rangle  &  =e^{-\frac{i}{\hbar}\hat{H}t}\left\vert
\Psi(t=0)\right\rangle \nonumber\\
&  =|\varphi_{+-}(t)\rangle\otimes|+-\rangle{}+|\varphi_{-+}(t)\rangle
\otimes|-+\rangle{}+|\varphi_{--}(t)\rangle\otimes|--\rangle{}+|\varphi
_{++}(t)\rangle\otimes|++\rangle\label{ROB-03}%
\end{align}
where we have used $|+-\rangle\equiv|+\rangle{}_{1}\otimes|-\rangle{}_{2}$,
$|-+\rangle\equiv|-\rangle{}_{1}\otimes|+\rangle{}_{2}$, $|--\rangle
\equiv|-\rangle{}_{1}\otimes|-\rangle{}_{2}$ and $|++\rangle\equiv|+\rangle
{}_{1}\otimes|+\rangle{}_{2}$. The last two terms represent the failure to
detect (no detector is ever perfect), and double detection (involving
something like a bounce, and corresponding to a small effect of order
$\lambda^{2}$).

One might think that the first two terms (the relevant ones for our
discussion) already show what one wants: we end up with the two alternatives
$|+-\rangle$ or $|-+\rangle$ breaking the symmetry, and we say that we just do
not know which one of the alternatives is selected by nature, or is
actualized. However, we will see that the situation is not that simple,
because the pair ($|+-\rangle$ , $|-+\rangle$) does not represent the only way
to characterize these alternatives.

\subsection{\label{TSP}The symmetry of the problem}

An inversion operator $\mathbb{\hat{P}}$ can be defined in such a way that it
changes $x$ by $-x$ in the wave function of the particle, and simultaneously
it interchange the states of the detectors, i.e.%
\[
\mathbb{\hat{P}}\left\vert \varphi\right\rangle \otimes\left\vert
\eta\right\rangle _{1}\otimes\left\vert \chi\right\rangle _{2}\equiv\left(
\hat{P}\left\vert \varphi\right\rangle \right)  \otimes\left\vert
\chi\right\rangle _{1}\otimes\left\vert \eta\right\rangle _{2}%
\]
where $\left\langle x\right\vert \hat{P}\left\vert \varphi\right\rangle
\equiv\left\langle -x|\varphi\right\rangle $.

It is easy to prove that $\mathbb{\hat{P}}$ is a symmetry of the Hamiltonian
of equation (\ref{ROB-01}), and that the vector of equation (\ref{ROB-02}),
representing the initial state of the composed system, is an eigenstate of
$\mathbb{\hat{P}}$, i.e.%
\[
\left[  \hat{H},\mathbb{\hat{P}}\right]  =0,\text{ \ \ \ \ \ \ }%
\mathbb{\hat{P}}\left\vert \Psi(t=0)\right\rangle =\left(  +1\right)
\left\vert \Psi(t=0)\right\rangle
\]
and therefore the value of $\mathbb{\hat{P}}$ is preserved by the time
evolution ($\mathbb{\hat{P}}\left\vert \Psi(t)\right\rangle =\left(
+1\right)  \left\vert \Psi(t)\right\rangle $). From this eigenvalue equation
and the definition of the operator $\mathbb{\hat{P}}$ we obtain%
\begin{align}
\hat{P}\left\vert \varphi_{--}(t)\right\rangle  &  =\left\vert \varphi
_{--}(t)\right\rangle \nonumber\\
\hat{P}\left\vert \varphi_{++}(t)\right\rangle  &  =\left\vert \varphi
_{++}(t)\right\rangle \nonumber\\
\hat{P}\left\vert \varphi_{-+}(t)\right\rangle  &  =\left\vert \varphi
_{+-}(t)\right\rangle \nonumber\\
\hat{P}\left\vert \varphi_{+-}(t)\right\rangle  &  =\left\vert \varphi
_{-+}(t)\right\rangle \label{ROB-04}%
\end{align}
The probabilities for the different possibilities of the two instruments
pointer states are%
\begin{align*}
\Pr\left(  +-\right)   &  =\left\langle \Psi(t)\right\vert \left\{  \hat
{I}_{P}\otimes\left\vert +-\right\rangle \left\langle +-\right\vert \right\}
\left\vert \Psi(t)\right\rangle =\left\langle \varphi_{+-}(t)|\varphi
_{+-}(t)\right\rangle \\
\Pr\left(  -+\right)   &  =\left\langle \Psi(t)\right\vert \left\{  \hat
{I}_{P}\otimes\left\vert -+\right\rangle \left\langle -+\right\vert \right\}
\left\vert \Psi(t)\right\rangle =\left\langle \varphi_{-+}(t)|\varphi
_{-+}(t)\right\rangle \\
\Pr\left(  ++\right)   &  =\left\langle \Psi(t)\right\vert \left\{  \hat
{I}_{P}\otimes\left\vert ++\right\rangle \left\langle ++\right\vert \right\}
\left\vert \Psi(t)\right\rangle =\left\langle \varphi_{++}(t)|\varphi
_{++}(t)\right\rangle \\
\Pr\left(  --\right)   &  =\left\langle \Psi(t)\right\vert \left\{  \hat
{I}_{P}\otimes\left\vert --\right\rangle \left\langle --\right\vert \right\}
\left\vert \Psi(t)\right\rangle =\left\langle \varphi_{--}(t)|\varphi
_{--}(t)\right\rangle
\end{align*}
Taking into account equation (\ref{ROB-04}), we obtain%
\begin{align*}
\Pr\left(  -+\right)   &  =\left\langle \varphi_{-+}(t)|\varphi_{-+}%
(t)\right\rangle =\left\langle \hat{P}\varphi_{+-}(t)|\hat{P}\varphi
_{+-}(t)\right\rangle \\
&  =\left\langle \varphi_{+-}(t)|\hat{P}^{2}\varphi_{+-}(t)\right\rangle
=\left\langle \varphi_{+-}(t)|\varphi_{+-}(t)\right\rangle =\Pr\left(
+-\right)
\end{align*}
As it was expected from the symmetry of the Hamiltonian and the initial
condition, the probability $\Pr\left(  +-\right)  $ to have only the
measurement instrument at $x=x_{0}$ excited is equal to the probability
$\Pr\left(  -+\right)  $ to have excited only the instrument at $x=-x_{0}$.

\subsection{Alternative choice of basis states}

Everything should be fine if one adopts an interpretation such as Bohr\'{}s,
in which the measurement instruments are classical objects, external to the
quantum theory. However, when the detectors are treated as quantum objects
themselves, things become more problematic: we will see that one seems to be
forced, not only to identify the quantum variables that are considered as
detectors and subject these to slightly different rules of treatment, but also
one would need to specify exactly how they are used. In other words, it seems
one should specify a-priori which variables are the appropriate ones we must
use in describing the situation. In the particular case we are dealing with
here, this issue can be easily illustrated.

In the previous subsections, we have used the vectors $\left\vert
+-\right\rangle $, $\left\vert -+\right\rangle $, $\left\vert --\right\rangle
$ and $\left\vert ++\right\rangle $ as a basis for the pointer states of the
two instruments.

But we might choose to work with the basis given by the following four vectors%

\begin{align}
|S\rangle &  \equiv\frac{1}{\sqrt{2}}\left(  \left\vert +-\right\rangle
+\left\vert -+\right\rangle \right) \nonumber\\
|A\rangle &  \equiv\frac{1}{\sqrt{2}}\left(  \left\vert +-\right\rangle
-\left\vert -+\right\rangle \right) \nonumber\\
|D\rangle &  \equiv\left\vert --\right\rangle \nonumber\\
|U\rangle &  \equiv\left\vert ++\right\rangle \label{ROB-05}%
\end{align}
This basis seems to be particularly convenient when discussing symmetry
related aspects of the problem. They can be used to expand the time dependent
state vector $\left\vert \Psi(t)\right\rangle =e^{-\frac{i}{\hbar}\hat{H}%
t}\left\vert \Psi(t=0)\right\rangle $ already obtained in equation
(\ref{ROB-03})%
\begin{equation}
\left\vert \Psi(t)\right\rangle =\left\vert \varphi_{S}(t)\right\rangle
\otimes\left\vert S\right\rangle +\left\vert \varphi_{A}(t)\right\rangle
\otimes\left\vert A\right\rangle +\left\vert \varphi_{--}(t)\right\rangle
\otimes\left\vert D\right\rangle +\left\vert \varphi_{++}(t)\right\rangle
\otimes\left\vert U\right\rangle \label{ROB-06}%
\end{equation}
where $\left\vert \varphi_{S}(t)\right\rangle \equiv\frac{1}{\sqrt{2}}\left\{
\left\vert \varphi_{+-}(t)\right\rangle +\left\vert \varphi_{-+}%
(t)\right\rangle \right\}  $ and $\left\vert \varphi_{A}(t)\right\rangle
\equiv\frac{1}{\sqrt{2}}\left\{  \left\vert \varphi_{+-}(t)\right\rangle
-\left\vert \varphi_{-+}(t)\right\rangle \right\}  $. The last equation
clearly exhibits the preservation of the initial symmetry.

Thus the question we must face is the following: Why would it be incorrect to
describe everything: the full Hilbert space, the evolution, including the
interaction of detectors with the particle using this last choice of basis. Is
there anything in the theory that would indicate which one is the correct
basis to talk about the problem?. Why is it that it seems less natural to use
the second rather than the first choice of basis? By the way, we note that
each one of the four elements of the full Hilbert space, appearing in the
above expression, are by themselves eigenstates of $\hat{P}$ with eigenvalue
$+1$.

\subsection{Decoherence}

One might object to the above discussion pointing out that only one dynamical
variable was considered for each measurement instrument, and it was the
variable associated with the pointer position, allowed to have only two
possibilities (excited and unexcited). This is clearly a highly idealized
representation. A real measurement instrument is a macroscopic object,
composed by an enormous amount of atoms. A more realistic description is to
consider the states of the instrument represented by vectors of a Hilbert
space which is the tensor product of a vector space associated with the
pointer variable and another vector space corresponding to an enormous number
of microscopic variables of the instrument, playing the role of what we might
call the environment.

Following the standard arguments of the theory of decoherence \cite{Zurek1981}
\cite{Zurek1982}, the interaction pointer-microscopic variables for the
instrument located at $x=+x_{0}$ may be described by the transformations%
\begin{equation}%
\begin{array}
[c]{ccccc}%
\left\vert -\right\rangle _{1}\left\vert \varepsilon_{-}\right\rangle
_{1}\longrightarrow\left\vert -\right\rangle _{1}\left\vert \varepsilon
_{-}\right\rangle _{1} &  & \left\vert +\right\rangle _{1}\left\vert
\varepsilon_{-}\right\rangle _{1}\longrightarrow\left\vert +\right\rangle
_{1}\left\vert \varepsilon_{+}\right\rangle _{1} &  & \left\langle
\varepsilon_{-}|\varepsilon_{+}\right\rangle _{1}\cong0
\end{array}
\label{DEC-01}%
\end{equation}
In this very rapid process, the two possible pointer states $\left\vert
-\right\rangle _{1}$ and $\left\vert +\right\rangle _{1}$ become correlated to
the approximately orthogonal environment states $\left\vert \varepsilon
_{-}\right\rangle _{1}$ and $\left\vert \varepsilon_{+}\right\rangle _{1}$.

Analogously, the interaction pointer-microscopic variables for the instrument
located at $x=-x_{0}$ gives%
\begin{equation}%
\begin{array}
[c]{ccccc}%
\left\vert -\right\rangle _{2}\left\vert \varepsilon_{-}\right\rangle
_{2}\longrightarrow\left\vert -\right\rangle _{2}\left\vert \varepsilon
_{-}\right\rangle _{2} &  & \left\vert +\right\rangle _{2}\left\vert
\varepsilon_{-}\right\rangle _{2}\longrightarrow\left\vert +\right\rangle
_{1}\left\vert \varepsilon_{+}\right\rangle _{1} &  & \left\langle
\varepsilon_{-}|\varepsilon_{+}\right\rangle _{2}\cong0
\end{array}
\label{DEC-02}%
\end{equation}
The interaction particle-instruments described in the previous subsections
produce the time dependent state described by equation (\ref{ROB-03}). If,
following the standard approach, this interaction is followed by the
interactions pointer-microscopic variables for both measurement instruments,
we obtain%
\begin{align}
\left\vert \Psi(t)\right\rangle  &  =\left\vert \varphi_{+-}(t)\right\rangle
\left\vert +-\right\rangle \left\vert \varepsilon_{+},\varepsilon
_{-}\right\rangle +\left\vert \varphi_{-+}(t)\right\rangle \left\vert
-+\right\rangle \left\vert \varepsilon_{-},\varepsilon_{+}\right\rangle
\nonumber\\
&  +\left\vert \varphi_{--}(t)\right\rangle \left\vert --\right\rangle
\left\vert \varepsilon_{-},\varepsilon_{-}\right\rangle +\left\vert
\varphi_{++}(t)\right\rangle \left\vert ++\right\rangle \left\vert
\varepsilon_{+},\varepsilon_{+}\right\rangle \label{DEC-03}%
\end{align}
where we used the notations $\left\vert \varepsilon_{\pm},\varepsilon_{\pm
}\right\rangle \equiv\left\vert \varepsilon_{\pm}\right\rangle _{1}%
\otimes\left\vert \varepsilon_{\pm}\right\rangle _{2}$ and omitted all tensor
product symbols $\otimes$ to produce a more compact expression.

Any observable involving only the pointer variables should have the form%
\begin{equation}
\hat{O}=\hat{I}_{P}\otimes\hat{O}_{pointers}\otimes\hat{I}_{E_{1}}\otimes
\hat{I}_{E_{2}} \label{DEC-04}%
\end{equation}
where $\hat{I}_{P}$ is the identity operator for the particle, and $\hat
{I}_{E_{1}}$ ($\hat{I}_{E_{2}}$) is the identity operator for the environment
of the instrument located at $+x_{0}$ ($-x_{0}$).

The mean value of the pointer operator (\ref{DEC-04}) in the decohered state
(\ref{DEC-03}) is%
\begin{align}
\left\langle \Psi(t)\right\vert \hat{O}\left\vert \Psi(t)\right\rangle  &
=\left\langle \varphi_{+-}(t)|\varphi_{+-}(t)\right\rangle \left\langle
+-\right\vert \hat{O}_{pointers}\left\vert +-\right\rangle +\left\langle
\varphi_{-+}(t)|\varphi_{-+}(t)\right\rangle \left\langle -+\right\vert
\hat{O}_{pointers}\left\vert -+\right\rangle \nonumber\\
&  +\left\langle \varphi_{--}(t)|\varphi_{--}(t)\right\rangle \left\langle
--\right\vert \hat{O}_{pointers}\left\vert --\right\rangle +\left\langle
\varphi_{++}(t)|\varphi_{++}(t)\right\rangle \left\langle ++\right\vert
\hat{O}_{pointers}\left\vert ++\right\rangle \label{DEC-05}%
\end{align}
We can now define an effective statistical operator in the space of the
pointer variables%
\begin{align}
\hat{\rho}_{pointers}  &  \equiv\left\langle \varphi_{+-}(t)|\varphi
_{+-}(t)\right\rangle \left\vert +-\right\rangle \left\langle +-\right\vert
+\left\langle \varphi_{-+}(t)|\varphi_{-+}(t)\right\rangle \left\vert
-+\right\rangle \left\langle -+\right\vert \nonumber\\
&  +\left\langle \varphi_{--}(t)|\varphi_{--}(t)\right\rangle \left\vert
--\right\rangle \left\langle --\right\vert +\left\langle \varphi
_{++}(t)|\varphi_{++}(t)\right\rangle \left\vert ++\right\rangle \left\langle
++\right\vert \label{DEC-06}%
\end{align}
This effective statistical operator can be used to compute the mean value of
equation (\ref{DEC-05}) in the space of the pointer variables%
\begin{equation}
\left\langle \Psi(t)\right\vert \hat{O}\left\vert \Psi(t)\right\rangle
=Tr\left(  \hat{\rho}_{pointers}\hat{O}_{pointers}\right)  \label{DEC-07}%
\end{equation}
The decoherence process has produced an effective state which is diagonal in
the basis $\{\left\vert +-\right\rangle ,\left\vert -+\right\rangle
,\left\vert --\right\rangle ,\left\vert ++\right\rangle \}$ for the pointer
states. However, in the present case we could not argue that this is
\textquotedblleft the\textquotedblright\ basis privileged by the decoherence
process. In subsection \ref{TSP}, we proved that $\left\langle \varphi
_{+-}(t)|\varphi_{+-}(t)\right\rangle =\left\langle \varphi_{-+}%
(t)|\varphi_{-+}(t)\right\rangle $, as a consequence of the symmetry of the
problem. Therefore the effective statistical operator $\hat{\rho}_{pointers}$
is also diagonal in any basis including two orthogonal linear combinations of
$\left\vert +-\right\rangle $ and $\left\vert -+\right\rangle $, together with
the vectors $\left\vert --\right\rangle $ and $\left\vert ++\right\rangle $.
One of these basis is the one defined by the vectors $\left\{  \left\vert
S\right\rangle ,\left\vert A\right\rangle ,\left\vert D\right\rangle
,\left\vert U\right\rangle \right\}  $ of equations (\ref{ROB-05}). Therefore,
the decoherence process on the symmetric problem we have considered is not
useful to privilege a basis of \textquotedblleft physical
states\textquotedblright, unless the two environments were initially in
different states. (See appendix for a theorem establishing the generality of
this problem).

\subsection{Predictability sieve criterion}

In the previous section, we studied the model of interest adding an
environment, according to the Zurek's recipe. We concluded that the
introduction of this environment is not enough to determine the actualization
basis of the pointer. The situation is the same as in Zurek et al. (1982)
\cite{Zurek1982} ,where the environment itself does not select the privileged
basis. According to Zurek, the determination of a privileged basis can be
carried out considering an additional criterion. This criterion is called
\textquotedblleft predictability sieve criterion\textquotedblright\ and
establishes that the privileged basis is given by the dominant term in the
Hamiltonian of the system. In the typical case, the system-environment
interaction Hamiltonian is dominant, hence the privileged basis will be the
eigenvector basis of the Hamiltonian of interaction $H_{int}$. For example, in
the model of Zurek
\[
H_{int}=\frac{1}{2}\left(  \left\vert \Uparrow\right\rangle \left\langle
\Uparrow\right\vert -\left\vert \Downarrow\right\rangle \left\langle
\Downarrow\right\vert \right)  \otimes\sum_{i=1}^{N}g_{i}\left(  \left\vert
\uparrow_{i}\right\rangle \left\langle \uparrow_{i}\right\vert -\left\vert
\downarrow_{i}\right\rangle \left\langle \downarrow_{i}\right\vert \right)
\]
where $\left\{  \left\vert \Uparrow\right\rangle ,\left\langle \Uparrow
\right\vert \right\}  $ are the eigenvectors of $S_{Z}$ for the system and
$\left\{  \left\vert \uparrow_{i}\right\rangle ,\left\langle \uparrow
_{i}\right\vert \right\}  $ are the eigenvectors of $S_{Z_{i}}$ for the
environment, the privileged basis are spin states with spin in $\hat{z}$
direction and the pointer indicates the spin in $\hat{z}$.

At this point we can not continue with generic analysis of the previous
section, where we studied the influence of a generic environment. To apply the
\textquotedblleft predictability sieve criterion\textquotedblright\ it is
necessary to clarify which is the environment Hamiltonian and which is the
interaction Hamiltonian. Following the steps of Zurek, we can choose an
interaction Hamiltonian. As an example we can specify that
\[
H_{int}=\left(  e_{++}\left\vert ++\right\rangle \left\langle ++\right\vert
+e_{+-}\left\vert +-\right\rangle \left\langle +-\right\vert +e_{-+}\left\vert
-+\right\rangle \left\langle -+\right\vert +e_{--}\left\vert --\right\rangle
\left\langle --\right\vert \right)  \otimes O_{E}
\]
where $O_{E}$ is some observable of the environment. Then, the eigenstates of
$H_{int}$ are%

\begin{align*}
H_{int}\left\vert ++\right\rangle \left\vert \varepsilon_{++}\right\rangle  &
=e_{++}\varepsilon_{++}\left\vert ++\right\rangle \left\vert \varepsilon
_{++}\right\rangle \\
H_{int}\left\vert +-\right\rangle \left\vert \varepsilon_{+-}\right\rangle  &
=e_{+-}\varepsilon_{+-}\left\vert +-\right\rangle \left\vert \varepsilon
_{+-}\right\rangle \\
H_{int}\left\vert -+\right\rangle \left\vert \varepsilon_{-+}\right\rangle  &
=e_{-+}\varepsilon_{-+}\left\vert -+\right\rangle \left\vert \varepsilon
_{-+}\right\rangle \\
H_{int}\left\vert --\right\rangle \left\vert \varepsilon_{--}\right\rangle  &
=e_{--}\varepsilon_{--}\left\vert --\right\rangle \left\vert \varepsilon
_{--}\right\rangle
\end{align*}
where $\left\vert \varepsilon_{\pm\pm}\right\rangle $\ are the eigenvectors of
$O_{E}$\ with eigenvalues $\varepsilon_{\pm\pm}$.

Thus, the privileged basis is the $H_{int}$ eigenstates basis, i.e. $\left\{
\left\vert +-\right\rangle ,\left\vert -+\right\rangle ,\left\vert
++\right\rangle ,\left\vert --\right\rangle \right\}  $ and the pointer
indicates the correct observable. Therefore, the problem seems solved because
the decoherence selects the possible states of the pointer in the proper way.
However, this method has two difficulties:

\begin{itemize}
\item The first is that it is necessary to introduce an interaction
Hamiltonian specially designed to obtain the desired results. If we choose a
different interaction Hamiltonian%
\[
H_{int}=\left(  e_{S}\left\vert S\right\rangle \left\langle S\right\vert
+e_{A}\left\vert A\right\rangle \left\langle A\right\vert +e_{D}\left\vert
D\right\rangle \left\langle D\right\vert +e_{U}\left\vert U\right\rangle
\left\langle U\right\vert \right)  \otimes O_{E}%
\]
the result is that the pointer is actualized in the basis $\left\{  \left\vert
S\right\rangle ,\left\vert A\right\rangle ,\left\vert D\right\rangle
,\left\vert U\right\rangle \right\}  $. Therefore, we must make the choice of
$H_{int}$ carefully, i.e. the introduction of the interaction Hamiltonian is
ad hoc.

\item Second, in the general case the introduction of such interaction
Hamiltonian breaks the symmetry of the total Hamiltonian. This is because the
Hamiltonian privileges one direction. In fact, if we permute 1 and 2 in
$H_{int}$ we have%
\begin{align*}
H_{int}\left\vert ++\right\rangle \left\vert \varepsilon_{++}\right\rangle  &
=e_{++}\varepsilon_{++}\left\vert ++\right\rangle \left\vert \varepsilon
_{++}\right\rangle \\
H_{int}\left\vert +-\right\rangle \left\vert \varepsilon_{+-}\right\rangle  &
=e_{-+}\varepsilon_{-+}\left\vert +-\right\rangle \left\vert \varepsilon
_{+-}\right\rangle \neq e_{+-}\varepsilon_{+-}\left\vert +-\right\rangle
\left\vert \varepsilon_{+-}\right\rangle \\
H_{int}\left\vert -+\right\rangle \left\vert \varepsilon_{-+}\right\rangle  &
=e_{+-}\varepsilon_{+-}\left\vert -+\right\rangle \left\vert \varepsilon
_{-+}\right\rangle \neq e_{-+}\varepsilon_{-+}\left\vert -+\right\rangle
\left\vert \varepsilon_{-+}\right\rangle \\
H_{int}\left\vert --\right\rangle \left\vert \varepsilon_{--}\right\rangle  &
=e_{--}\varepsilon_{--}\left\vert --\right\rangle \left\vert \varepsilon
_{--}\right\rangle
\end{align*}
the only case where the symmetry is not broken is $e_{-+}\varepsilon
_{-+}=e_{+-}\varepsilon_{+-}$. But if we take this case we have degeneration,
thus any lineal combination of $\left\{  \left\vert +-\right\rangle
,\left\vert -+\right\rangle ,\left\vert ++\right\rangle ,\left\vert
--\right\rangle \right\}  $ is an eigenstate of $H_{int}$, therefore $\left\{
\left\vert S\right\rangle ,\left\vert A\right\rangle ,\left\vert
D\right\rangle ,\left\vert U\right\rangle \right\}  $ is other $H_{int}$
eigenstates basis. In the present case, the predictability sieve criterion can
not select univocally a preferred basis, and, in particular, can not be used
to identify what we might intuitively feel is the correct one.
\end{itemize}

The theorem we present in the appendix ensures that we will face this issue in
the cosmological problem at hand. Thus we conclude that, even taking into
account the predictability sieve criterion, the approach based just on
decoherence is not helpful in offering a solution to our predicament.

\section{ Addressing the problem in the various interpretational schemes}

One of the most clear evidences of the persistent state of confusion about
quantum theory is the existence of a plethora of interpretations. Nothing like
this happens with the other physical theories. There is no pressing/critical
questions about the interpretation of Maxwell electrodynamics, or that of
Einstein's theories of Relativity, either Special or General (see however
\cite{Barbour}).

An exhaustive analysis of each one of those interpretations, or a detailed
comparative study of their relative advantages and disadvantages is clearly
outside the scope of the present manuscript. However, we will briefly survey
the field in order to show that, in facing the problem that concerns us here,
they all seem to come short. Before embarking in a more detailed way on that
path, let us give the definitions of some concepts that will be used
frequently in what follows.

Given a quantum theoretical description of a problem, we assume that one is
given a Hilbert space, a Hamiltonian, and the set of observables. However, one
often wants to demand that the discussion be carried out in a certain basis of
the Hilbert space. Such a choice of preselected and privileged basis is called
a context, and it often dictates essential aspects of the interpretation such
as \textquotedblleft collapse or actualization\textquotedblright. How is that
choice made, will, in general, depend on the particular interpretative scheme
one wants to employ. Let us recall that selecting a context is equivalent to
choosing an orthogonal basis of the Hilbert space, and requiring that all
vectors and operators be described in such basis whenever the interpretation
of the mathematics is required. In this setting, the coefficients of the
corresponding expansions are then taken as yielding the corresponding
probabilities. The concept associated with this is that of \textquotedblleft
actualization\textquotedblright\ or that of some alternative notion of
\textquotedblleft a possibility becoming actual\textquotedblright. The precise
meaning of the word \textquotedblleft actual" naturally depends on the type of
interpretation. In the Bohm De Broglie interpretation (which is often
considered as involving hidden variables) the actualization is permanent, as
it refers to the value of the hidden variable representing the
\textquotedblleft particle's position\textquotedblright\ ---or, more
generally, the point in configuration-space that together with the wave
function represents the physical situation--- corresponds to the actualized
value of the position $\vec{x}$. In the other cases, the notion of
actualization is associated with a change in the state of the system that,
depending on the interpretation, is brought about by various causes and has
different connotations.

It is well known that the logic of quantum mechanics is not a Boolean logic
but a quantum logic \cite{QLogic}. Our brain knows how to reason with Boolean
logic, but it is unable to use quantum logic (at least at the present time).
Then, in applying the theory, we must somehow combine quantum mechanics with
Boolean logic. Moreover, we can consider the problem in just a particular
instant for some instantaneous type of interpretation or consider periods of
time within one of the historical interpretations. There are interpretations
that consider a special role for the apparatuses, often taken as classical and
outside the scope of the theory, and consider that the theory only pertains to
the results of the usage of these apparatuses to study a system. In some of
them, the process of measurement produces \textquotedblleft the collapse of
the wave function\textquotedblright, e.g. in the Copenhagen interpretation
(see \cite{SEP:Copenhagen}). In one extreme we find the approaches where the
posture is that \textquotedblleft the system does not even
exists\textquotedblright\ in the same physical sense as the apparatuses and
observers. In the realistic interpretations, the measureing apparatus are
missing, or not considered as essential, and they are substituted by an
actualization of the wave function (e.g. in the modal interpretations, see
\cite{SEP-Modal}).

Here, we will discuss how the most popular interpretations deal with the
simplified version of Mott's problem and with the cosmological problem that
motivates our analysis.

\subsection{Classical apparatuses Interpretation}

In this interpretation, there is a coexistence between the classical world and
the quantum world. The context (i.e. the basis of the Hilbert space in which
one analyzes the situation) is determined by the classical measuring apparatus
( one assumes that these are clearly specified ). This interpretation is
supposed to be the interpretation \textquotedblleft for all
weathers\textquotedblright. Accordingly, one is supposed to take the view that
the only things that truly exists are those measuring apparatus, including, of
course, the preparation apparatuses which are just other kind of measuring
apparatuses. Those are taken to be always macroscopic, and, therefore, so the
posture states, they must be treated classically with the usual boolean
classical interpretation. The rest of the formalism is just a mathematical
characterization of a microscopic world, but not a realistic description
thereof, something that, in any event, is seen as lying outside the realm of
science, and thus it is not considered as corresponding to one which could be
taken as having physical reality. This seems to be the way in which many
experimental physicists have learned to think, and which ensures that they
never make mistakes. In this interpretation, the fundamental requirement is
the existence of a clearly identified classical measuring apparatus, the
description of which lies outside of the scope of the quantum theory.

Here, one would have to say that the description of the detectors at the
quantum level is simply inappropriate. The detectors must be regarded as
macroscopic, and, thus, intrinsically classical systems, and the classical
states are, therefore, those which would tie them with the first basis. I.e.
they are excited or non excited (but of course, being classical they are not
described in the language of state vectors or operators in any Hilbert state).
The problem, of course, is that such posture violates the rules we have set up
to ourselves to solve the problem only within a scheme that would be
applicable to the cosmological problem that motivated our analysis in the
first place. The point is that, in that situation there are simply no systems
that can be envisioned as playing the role of the measuring apparatuses.
Apparatuses will emerge only after complex measuring instruments would be
designed and built by sapient beings. And complex instruments and beings
require the existence of planets, stars and generally, inhomogeneous and
anisotropic regions to live and evolve, and, thus, this view is simply
unsuitable to address the cosmological problem.

\subsection{Copenhagen Interpretation}

This interpretation is accepted as the official one \cite{SEP:Copenhagen}. The
vast majority of the books are written based on its rules and concepts. Here,
measurements are viewed as forcing the quantum collapse of the state of the
system into the eigenvector determined by the measurement and its outcome. The
collapse is associated with measurement. According with the textbook by Cohen
Tanuodji et al. \cite{Cohen} pp. 221, the collapse postulate is the following:
\textquotedblleft If the measurement of a physical quantity $A$ on the system
of state gives the result $a_{i}$, the state of the system, immediately after
the measurement is the projection onto the subspace associated with
$\left\vert a_{i}\right\rangle $\textquotedblright. This interpretation takes
the view that, even if the apparatuses might be described at a quantum level,
there are distinct physical processes called measurements, which are governed
by very special rules: When a measurement takes place, the state of the system
undergoes a sudden jump into one of the eigenvalues of the observable being
measured and the probability for such jump is given by the Bohr's rule. This
interpretation, thus, involves the notion of measurement as an independent
concept, or in some presentations, as lying outside the scope of theory: It is
something that can not be described in terms of the other concepts of the
theory: states and/or operators on a Hilbert space. This is, in a sense, the
most widely used interpretation, is presented in most text books, and has been
subjected to multiple criticisms (see \cite{Ballentine} for criticisms). In
this interpretation, the essential component, the existence of which is taken
for granted, is an external measuring device, a quantum system, which somehow
produces/ induces the collapse.

In this case, one takes the measurements as triggering the quantum collapse of
the state of the system. Thus Mini-Mott problem would be solved by describing
the state of the system (now the particle and detectors) using the basis which
is appropriate to describe the measurement. The point is that the measurement
would have to be described by something that goes beyond the mere
identification of the interaction hamiltonian, because, as we have seen, we
can describe it in either, the symmetric or the non-symmetric basis. In the
case of Mini-Mott, we would have to say that, the detectors are somehow
constructed to detect the particle either at one position or the other, and
this characterization can not be made simply by writing the interaction
hamiltonian. The measurement is identified as a special type of interaction
that is subject to spacial rules that do not apply to all interactions. The
problem again is that this kind of solution would not be applicable to the
cosmological problem, as in that situation there are no measuring apparatuses,
and no measurements. As before, measurements require complex beings and those
require planets, stars and generally inhomogeneous and anisotropic conditions
to emerge. Thus, unless one want to invoke some God-like entity predating the
emergence of structure in our universe, and which can perform measurements, we
must acknowledge that, there is, within this interpretation, no solution to
our cosmological problem.

In short, the problem with the instrumentalist interpretations, such as the
previous two, is that in the situations at hand, there are simply no
instruments and no observers (recall we are dealing with the inflationary
regime and the process of generation of inhomogeneities and anisotropies that
would eventually evolve into galaxies and stars that can in turn be the
regions where life, intelligence, and even instruments can arise). In fact, it
is one of the goals of cosmology to provide an explanation of the emergence of
those conditions which lead to the generation in our universes of structures
such as galaxies, planets and eventually living organisms such as ourselves,
capable of making observations, and building instruments. Therefore, the
instrumentalist path seems to be closed to us, at least in as much as we are
focussing on the cosmological problem and on the related ones such as the
Mini-Mott problem.

\subsection{Statistical Interpretation}

In this interpretation of quantum mechanics, the quantum state is interpreted
as an abstract quantity that characterizes the probability distribution for an
ensemble of identically prepared systems. That is, ensembles, and not
individuals systems are considered as central to the theory, i. e. idealized
sets containing infinite copies of identical systems. The quantum state
corresponds to a collective description of all elements of the ensemble but
not of each individual element. A quantum state corresponding to a
superposition of different macroscopic states,is not seen as constituting any
problem within this approach. It is just taken to represent a potential set of
results and not the coexistence thereof. Thus, the main element to which the
theory applies is the statistical ensemble,and not the individual system. We
note that the application of the formalism within this interpretation, to any
specific situation, requires the identification of a context (i.e. the basis
of the Hilbert space in which one analyzes and discusses the situation) In the
case where the experiment under consideration involves the measurement of a
property, the context is determined by that property, and thus indirectly by
the measuring devices. However, when there are no measuring devices
identified, the interpretation often presupposes some choice of a preferred
context. Without the context, one does not know what exactly is the ensemble
one wishes to discuss, and, in particular, one does not know how should, the
individual elements of the ensemble, be characterized.

Within this interpretation one considers that there must be an ensemble of
copies of the system and that the individual systems in the ensemble become
actualized in the various possibilities. However, as we have indicated
previously, this interpretation requires a selection of the privileged basis
to talk about the corresponding statistics. In the absence of measuring
devices and/or anything that can play the role of observers, we have no way of
doing so. Exactly in the same way that we could not argue convincingly that we
should choose, for instance, the first over the second basis in Mini-Mott problem.

Furthermore, let's recall that, according to the statistical interpretation,
one must adopt the position that quantum theory does not describe individual
systems, (in this case, the Universe), but only some statistical ensembles of
similarly prepared systems. This raises an important point. In order to make
statistics over an ensemble one needs to be able to talk about each individual
system that makes up the ensemble. Statistical averages of quantities are
defined in terms of the individual values those quantities takes on each
element of the ensemble. If one takes the view that individual systems are
described in classical terms, one immediately faces the problem of having, in
principle, the possibility of assigning to each individual system values of
quantum incompatible observables. If, alternatively, one invokes a quantum
characterization of individual systems, one must face the problem of having to
talk about the measurement problem or some counterpart thereof. In particular,
if we want to consider the statistical characterization of each system one
must face the choice of basis or context one will use to talk about them.

Thus, we are faced again with the issue: how are we to distinguish
\textquotedblleft a measurement" from other interactions? If we presuppose
that there are macroscopic variables that are accessible to us as observers,
as done in Ballentine's book, one is, of course, bringing the observer into
the picture as the means to make the selection of the privileged basis.
However, \textquotedblleft macroscopic" and \textquotedblleft accessible" are
clearly words that have a deep anthropocentric connotation.

In the case of our Mini-Mott example, one would need, not only to identify the
detectors as playing the role of measuring apparatus, but one would have to
postulate the appropriate basis to talk about the system (or at least about
the detectors) and use the statistical interpretation which is deemed to be
the natural one for macroscopical and accessible variables of the apparatus.
The choice which seems natural to us, given our experiences, does not seem to
be indicated by anything present in the theory\footnote{On the other hand, it
is worth noting that the Hamiltonian of interaction between particle and
detector has a explicitly local form in the first basis but not in the second.
This might be used but it would have to be explicitly formulated as part of
the theory. Spontaneous localization theories, and de-Broglie Bohm approaches,
for instance focus on position as playing a preferential role.}.

In fact, we must wonder why is it that the second basis might not be
considered as tied with an accessible and macroscopic variable. It seems we
must argue that the values \textquotedblleft symm\textquotedblright\ and
\textquotedblleft anti\textquotedblright\ are for some reason, not accessible,
but the theory does not tell us why. This suggests that there is something
that escapes Quantum theory and needs to be understood at some deeper level.
Moreover, even if the exact nature of the ensemble were fully determined, it
seems clear that the appropriate context is not identified and unambiguously
selected by the theory. The fact is that the statistical interpretation lacks
a clear criterion for such context selection.

\subsection{Modal Interpretations}

The name of these interpretations comes from the fact that in the early
versions they were related to a certain type of Modal logic proposed by van
Frasen \cite{FRASEN}. According with the Stanford Encyclopedia of Philosophy,
the general features of modal interpretations are:

\begin{itemize}
\item The interpretation is based on the standard formalism of quantum
mechanics, with one exception: the projection postulate is left out.

\item The interpretation is realist, in the sense that it assumes that quantum
systems possess definite properties at all instants of time.

\item Quantum mechanics is taken to be fundamental: it applies both to
microscopic and macroscopic systems.

\item The dynamical state of the system (pure or mixed) tells us what the
possible properties of the system and their corresponding probabilities are.
This is achieved by a precise mathematical rule that specifies a probabilistic
relationship between the dynamical state and possible value states.

\item A quantum measurement is an ordinary physical interaction. There is no
collapse of the dynamical state: the dynamical state always evolves unitarily
according to the Schr\"{o}dinger equation.
\end{itemize}

These are interpretations that do not depend on the existence of instruments
or observers as differentiated objects outside the quantum theory. These
interpretations replace the postulate of collapse by one of actualization.
There is a privileged context in which system properties take definite
values{}{}. The difference between them is that they choose different contexts.

One of the appealing features of the modal interpretations is that there is no
collapse, and the evolution is always unitary.
In our example this means that the symmetric initial state evolves with a
symmetric Hamiltonian, and then the symmetry will be present in the state at
all times. This is a general theorem\footnote{Let $\hat{S}$ be a symmetry
operator and $|\Psi(0)\rangle$ an initial symmetric state, i.e. $\hat{S}%
|\Psi(0)\rangle=|\Psi(0)\rangle$. Let $\hat{H}(t)$ be the system's
hamiltonian, taken to be invariant under the symmetry i.e. $[\hat{H}%
(t),\hat{S}]=0$. Then $\hat{S}|\Psi(t)\rangle=\hat{S}e^{i\int_{0}%
^{t}H(t^{\prime})dt^{\prime}}|\Psi(0)\rangle=e^{i\int_{0}^{t}H(t^{\prime
})dt^{\prime}}\hat{S}|\Psi(0)\rangle=e^{i\int_{0}^{t}H(t^{\prime})dt^{\prime}%
}|\Psi(0)\rangle=|\Psi(t)\rangle$ i.e. the evolved state is also symmetric.}
and, as has been shown explicitly in Section \ref{TSP}\ , the state that
results from the evolution never loses their symmetry. One fundamental issue
that the modal interpretations would have to face, in attempting to address
the problem at hand, is the following: If the context selected by the
particular modal interpretation is such that includes, as an element of the
preferred basis, a state that has the property of homogeneity and isotropy
such as the initial state of Mini-Mott, or the Bunch Davies state of the
quantum field in inflationary cosmology, why would that cease to be the case
at all other (later) times. I.e. why would the symmetric states cease to be
part of the preferred basis at later times. This issue seems to be an
inescapable one, because, unless such change of context takes place, one could
not explain the breaking of the symmetry. Additionally, some of the modal
interpretations that we present have \textquotedblleft no-go\textquotedblright%
\ theorems \cite{NoGoModal} that make them unsuitable for the general
interpretation of the theory.

\subsubsection{Atomic modal interpretation}

This interpretation assumes that there is, in nature, a fixed set of mutually
disjoint atomic quantum subsystems that constitute the building blocks of all
the global quantum systems. i. e. it establishes a preferred factorization of
the Hilbert space \cite{ModalAtomic}. It decomposes the system (called
molecular) in \textquotedblleft atomic\textquotedblright\ blocks $\left\{
\alpha_{q}\right\}  $, each in a state $\hat{\rho}_{q}$, now the privileged
base is $\left\{  \left\vert iq\right\rangle \right\}  $. The reduced state of
each block is%
\begin{equation}
\hat{\rho}_{q}=\sum_{i}\rho_{i}^{(q)}\left\vert iq\right\rangle \left\langle
iq\right\vert
\end{equation}
Thus, one can set properties on the subsystems of the total system. However,
the basis is undetermined in each subsystem.

The main problem for the Atomic modal interpretation is to justify the
assumption that there is a preferred partition of the universe, and to provide
some idea about what this factorization should look like. Moreover we
generally do not end up with a well specified basis for the complete system.
In other words, the shortcomings of this interpretation are intimately
connected with some the the issues we must resolve in our quest to address the
Mini Mott or the Cosmological problems : The choice of the preferred basis.

\subsubsection{Biorthogonal-decomposition modal interpretation}

This interpretation sometimes is known as \textquotedblleft Kochen-Dieks modal
interpretation\textquotedblright\ \cite{ModalSmith}. The definite-valued
observables are picked out by the biorthogonal (Schmidt) decomposition of the
pure quantum state of the system separated into subsystems. The state is
decomposed into Schmidt basis%
\[
\left\vert \psi^{\alpha\beta}\right\rangle =\sum_{j}c_{j}\left\vert
c_{j}^{\alpha}\right\rangle \otimes\left\vert c_{j}^{\beta}\right\rangle
\]
And the states $\left\vert c_{j}^{\alpha}\right\rangle $ $\left\vert
c_{j}^{\beta}\right\rangle $\ define the properties that take a defined value.
This interpretation has the obvious difficulty that a system can be decomposed
into subsystems in a variety of different ways.

However, the fact that a system can be decomposed in a variety of different
ways leads to multiple alternative choices for the biorthogonal decomposition
and, thus, introduces the following problem \cite{SEP-Modal}: In order to
apply this interpretation, we need to know in advance what is the privileged
basis (or decomposition). Again, the problem with this interpretation, is the
problem we want to solve. Many authors believe that those problems can be
solved by appealing to quantum decoherence. But as we have seen, in the
situations under consideration here, decoherence simply can't not perform the
task one expects from it.

\subsubsection{Perspectival modal interpretation}

In this interpretation the properties of a physical system have a relational
character and are defined with respect to another physical system that serves
as a \textquotedblleft reference system\textquotedblright%
\ \cite{ModalPerspectival}. The starting point is that the universe is in a
pure state $\left\vert \psi\right\rangle \left\langle \psi\right\vert $,
evolves according to the Schr\"{o}dinger equation and never collapses. Using
the partial trace it is possible to compute the state of a subsystem
\textit{S} with respect to the rest of the\textbf{ } universe
\[
\rho_{U}^{S}=Tr_{\left(  U\backslash S\right)  }\left\vert \psi\right\rangle
\left\langle \psi\right\vert
\]
the spectral resolution of $\rho_{U}^{S}$ defines the properties that take a
defined value.

As in shown in our analysis of the role of decoherence in the situations at
hand, here we need to divide the whole system into relevant sub-system and
environment, and, given the symmetry of the situation it is impossible to find
the privileged basis (See the appendix for a theorem exhibiting the generality
of the problem). In fact, if we were to admit different partitions between
system and environment we could face logical contradictions \cite{looming}
\cite{looming2}.

\subsubsection{Modal-Hamiltonian interpretation}

According to this interpretation, a quantum system $S$ is represented by a
pair $(\mathcal{O},\hat{H})$ where (i) $\mathcal{O}$ is a space of all
possible operators, (ii) $\hat{H}\in\mathcal{O}$ is the time-independent
Hamiltonian of the system $S$, and (iii) the state evolves according to the
Schr\"{o}dinger equation \cite{ModalHamiltoniana}. Given a quantum system $S$,
the actual-valued observables of $S$ are the Hamiltonian $\hat{H}$, and all
the observables commuting with $\hat{H}$ and having, at least, the same
symmetries as $\hat{H}$. This interpretation is particularly suitable to be
applied to closed systems where there is a no time-dependent Hamiltonian. But
it cannot be used in the general case.

In fact, for this particular modal interpretation, one of it's axioms prevents
its application to systems with truly time dependent Hamiltonians. In general,
one assumes that,if one has a time dependent Hamiltonian, it is because one
has failed to consider the complete system, and that what one has been
considering is just a part of a larger system with a time independent
Hamiltonian. In other words, in order to apply the formalism one has to
correctly identify the complete system, having a time independent Hamiltonian.
While there is some progress in the attempts to apply this interpretation in
quantum field theory, the proponents have never considered a curved spacetime
\cite{MHIfields}. However, we must stress that, in a general relativistic
setting there is, in general, no such time independent Hamiltonian: In fact,
if one includes gravity, the full Hamiltonian vanishes, and when one restricts
consideration to the matter sector alone, the Hamiltonian depends on arbitrary
choices of lapse and shift functions. In the cosmological context at hand, the
matter Hamiltonian depends on time due to the expansion of the universe.

\subsection{de Broglie-Bohm interpretation}

One of the many perspectives offering an interpretation to the results of the
quantum mechanics experiments can be traced back to L. de Broglie, and was
resurrected and refined by David Bohm. In the work \cite{Bohm1952a}
\cite{Bohm1952b}, Bohm wrote the Schr\"{o}dinger equation in a particular way:
He separated the module $R$ and phase $S$ of the wave function obtaining a set
of coupled equations governing the evolution of $R$ and $S$. One of these
equations is easily interpreted as a probability conservation equation upon
the introduction of an appropriate ensemble of particles and an equation
determining each particle's velocity in terms of the gradient of the phase $S$
at the particle's instantaneous position. The other equation is formally
identical to the Hamilton-Jacobi equation of classical mechanics, where
classical potential is added to the quantum terms. This novel terms are then
interpreted as a quantum contribution to the potential. This perspective
indicates that the phase $S$ can be interpreted as the generating function,
which allows the calculation of the possible trajectories of the particle.
Thus one obtains a deterministic quantum mechanics in which the particles have
definite positions and velocities at any given time, i.e. in this approach
particles do have well defined paths. In fact, this approach is sometimes
considered, not just as an interpretation of quantum mechanics, but as a
different theory, and under certain circumstances (see \cite{Valentini}) one
can expect predictions that differ from standard quantum theory.

These ideas have been discussed and refined, and a more complete version can
be found in the book \textquotedblleft Quantum Theory of
Motion\textquotedblright\ by Peter R. Holland \cite{Holland}. Within this
scheme, the trajectory of a quantum particle is well defined once its initial
condition is determined as in the classic case. For example, in the
double-slit experiment, the quantum potential has the shape of
\textquotedblleft gutters\textquotedblright. The gutters start at the particle
gun and end at the screen. The potential is such that the density of
\textquotedblleft gutters\textquotedblright is higher in regions where, from
the orthodox view, constructive interference is expected, and is smaller in
destructive interference areas. In this experiment, for reasons of practical
nature, it is impossible to determine which of the gutters will be taken by a
particle which departs from the gun. However, if we observe the arrival point
of the particle on the screen, it is possible to determine which gutter was
taken and, consequently, its initial condition.

Bohm's theory as it exists today is not directly applicable to solve our
proposed problem in a completely satisfactory way. That approach is tied
intrinsically with particle quantum mechanics, as it involves, in a sense, a
choice of a preferential basis, through the fact that the theory singles out
the particle's position variable, as the one that is permanently actualized.
In any attempt to apply this approach to the cosmological problem at hand, the
first thing we would need to do is to select the corresponding special
variable for the case of a field theory. One might be inclined to take the
field amplitude as playing such spacial role, by arguing, for instance, that ,
in general, such role must be assigned to be the configuration variables. The
issue would become more delicate and problematic, if what we have is a gauge
field theory. In any event, it is clear that some nontrivial choices must be
made. Once such choices are made, one might apply the field theoretical
version of the d' Brogile-Bohm approach ,to the cosmological problem. This is
in fact, a path taken in the works given in reference \cite{CosmoBohm}.

The second point we should make is that in this scheme, the initial conditions
involve not only the initial wave function of the system, but also the initial
value of the configuration variables (i.e the particle's position in the case
of non-relativistic quantum mechanics). In that sense, in considering the
question of symmetry of the initial conditions, one must consider both
aspects. In fact, given any system, the key of the explanation of its behavior
is to be found in the initial condition for the configuration variables.

For instance if we study our problem from the Bohmian perspective, it is clear
that the symmetry of the arrangement, and the initial wave function will
induce a symmetric quantum potential. However, the initial condition for an
individual particle should, according to the spirit of the approach, be chosen
in a random fashion from an appropriate \textquotedblleft equilibrium
distribution". Generically, such initial condition will NOT be symmetric. This
would seem to account for the fact that, in practice, we observe that the
particle (e. g. in the mini-Mott example) is detected only by one of the
detectors. In this sense, the path of the particle was determined by the
initial condition, and was predetermined from the beginning. This feature,
thus exhibits the fact that the initial condition was not symmetrical. I. e.
in this theory, the asymmetry that we intend to deduce must be taken as
introduced from the start. Bohm offered a replied to a critique related to
this point which can be found in \cite{BohmDirac}. There he argued that, in
any given experiment, the particle interaction with the environment would push
the particle to one side or the other. This argument is valid, of course, in
any realistic experimental situation, but we can not apply it to the
cosmological case because, by definition, the universe as a whole, has no
environment. Moreover, as we have seen already, the introduction of an
environment, which is subject to a quantum mechanical treatment, and which is
assumed to share the symmetry properties of the problem, gives rise to the
same problems which appeared in the discussion of the decoherence perspective.
Thus, in a strict dBB approach to the problem, the origin of asymmetry would
be found in the initial conditions of the \textquotedblleft hidden
variable\textquotedblright\ of the theory, (i.e. the position, or in general
something like the configuration variable) just as in the decoherence proposal
it must be associated with some asymmetry in the state of the environment. Our
argument is supported by the fact that all the work in cosmology which was
done under this perspective introduces the asymmetry in the initial condition
of the universe \cite{CosmoBohm}. For this reason we must consider that the
d'Broglie- Bohm approach can not be said to offer a satisfactory explanation
of the emergence of asymmetry: Simply speaking, the asymmetry is there from
the start.

\subsection{Many worlds interpretation}

The fundamental idea of this interpretation is that, in addition to the world
that we can see, there are many parallel worlds that make up the totality of
what exists \cite{SEP-ManyWorlds}. At every time a quantum experiment
involving different potential outcomes with non-zero probability is performed,
all outcomes are actually obtained, each in a different world, even if we are
aware only of the world with the outcome we have seen. In fact, quantum
experiments take place everywhere and very often, not just in physics
laboratories: even the irregular blinking of an old fluorescent bulb is a
quantum experiment \cite{SEP-ManyWorlds}. In Everett words:

\textquotedblleft We thus arrive at the following picture: Throughout all of a
sequence of observation processes there is only one physical system
representing the observer, yet there is not a single and unique state of the
observer (i.e. a collapse state)(which follows from the representation of the
interaction-system). Nevertheless, there is a representation in terms of
superposition, each element of which contains a definite observer state and a
corresponding system state (i.e. the systems ordinary state). Thus with each
succeeding observation (or interaction) the observer state `branches' into a
number of different states. Each branch represents a different outcome of the
measurement and the corresponding eigenstate of the object-system state. All
branches exist simultaneously in the superposition after any given sequence of
observations. Thus if we represent the free evolution of the system as
bifurcating paths at actualization points from which the multiple alternatives
emerge, to bifurcate again and again offering a multiplicity of worlds, we end
up with the multi-temporal object to which this interpretation refers. By the
way each one of the vertices or bifurcations points must be associated with a
corresponding context that determines the bifurcation basis at that event. It
is amazing how much can be said about such bizarre picture of
reality.\textquotedblright\ See \cite{Everett1957} pp. 459

In this interpretation, one of the problems one must face is that it is not
specified on what basis the bifurcation occurs, or exactly when, and under
which conditions it takes place. Thus, again, it becomes necessary to specify,
among other things, a preferred basis, or context. According to some authors,
this basis can be obtained from a decoherence type of analysis
\cite{SEP-Decoherence}. As the many worlds approach requires a privileged
basis which characterizes the branchings, it is clear that in the case of the
Mini-Mott problem we would have to determine if the branchings are to be
associated, for instance, with the first or second decomposition of the state,
i.e the expression given in eq. (\ref{ROB-03}), or rather the one in eq.
(\ref{ROB-06}). There seems to be simply nothing at all, intrinsic to the
setting, to help us determine how should this selection must be made. It is
only if we invoke something like an observer with a conscious brain which for
its own intrinsic reasons is entangled in a particular simple (diagonal) way,
that one might be able to argue that the description of the detectors of our
Mini Mott situation should be made in the first rather than the second basis.

\subsection{Interpretations based on histories}

Interpretations based on histories consider a formalism suitable to give
descriptions of quantum systems involving properties at different times. We
know two versions:

\begin{enumerate}
\item[a.] \textit{Consistent Histories} \cite{HistConsistentes}: The theory of
consistent histories is a framework to consider, using a quantum language, the
properties of a quantum system at different times. It deals with a series of
times $\{t_{i}\}_{i\in I}$ and the specification at each $t_{i}$ of a
decomposition of the Hilbert space of the system into suitably chosen
subspaces. The choice of one such subspace at each $t_{i}$ is known as a
\textquotedblleft coarse grained history". The resulting set of coarse grained
histories is called a realm if the quantum mechanical amplitude for
interference between two coarse grained histories in the set vanish.

Under such circumstances, the scheme assigns probabilities to each coarse
grained history in a manner that would correspond to Born's rule. The point,
however, is that only when the consistency condition is satisfied for the set
of histories, can it be regarded as proving a consistent characterization of
the system's development. This is the origin of the name \textquotedblleft
interpretation of consistent histories\textquotedblright.

More specifically, the scheme is based on the consideration, given a quantum
state of the system represented by a density matrix $\hat{\rho}$ at time
$t_{0}$, of families of histories characterized by a set of projection
operators $\{\hat{P}_{n}(t_{n})\}$, each of which is associated with the
system possessing a value of certain physical property in a given range at a
given time. A family $F$ of such projectors is called self consistent, if the
resulting histories do not interfere among themselves. Then, the scheme
assigns probabilities to each individual \textquotedblleft\ coarse grained
history\textquotedblright\ within the family according to the rule:%
\begin{equation}
P=Tr(\hat{P}_{n}(t_{n})U(t_{n},t_{n-1})\hat{P}_{n-1}U(t_{n-1},t_{n-2}%
)......\hat{P}_{2}U(t_{2},t_{1})\hat{P}_{1}U(t_{1},t_{0})\hat{\rho}%
U(t_{n},t_{0})^{\dagger})\label{ProbCH}%
\end{equation}

where the $U$'s stand for the standard unitary evolution operators connecting
two times. The fact, however, is that, in general, there exists a multiplicity
of possible choices of the realm, and the scheme does not indicate, at
fundamental level, which one is to be used in each circumstance.

\item[b.] \textit{Generalized contexts formalism} \cite{HistContextuales}: It
is also a formalism suitable to give descriptions of quantum systems involving
properties at different times. It is, in a sense, a refinement of the previous interpretation.

Again, the interpretation deals with a time series $\{t_{i}\}_{i\in I}$ and
the specification at each $t_{i}$ of a characterization of the system by a
suitably chosen set of properties. Each of the possible selections of
properties of a quantum system is called a "generalized context" or
\textquotedblleft description".

The quantum properties of a generalized context should satisfy two types of
compatibility conditions:

1) For properties at the same time, the corresponding projection operators
should be generated by a projective decomposition of the identity operator,
i.e. by a collection of mutually orthogonal projectors adding up to the
identity operator.

2) For properties at different times, the corresponding projectors should
commute when translated to a common time.

This formalism was successfully tested to give suitable descriptions of a
measurement process \cite{mp}, the double slit experiment with and without
measurement instruments \cite{pr}, and the quantum decay process \cite{dp}.

This formalism corresponds to a modification of the consistent histories
approach, having the advantage that some problematic histories that can arise
in the consistent theory are eliminated. Moreover, in contrast with the
former, here, the compatibility conditions for properties at different times
are state independent, and therefore the allowed generalized contexts giving
descriptions of a quantum system are also state independent. This is an
interesting feature, because in the usual axiomatic theories of quantum
mechanics the state is considered as a functional on the space of observables,
and it enters into the theory in a somehow subordinate position.
\end{enumerate}

As we indicated, the main problem with this type of approach is that, although
the scheme seems satisfactory, one has selected a particular decoherent
family, and there exists, in principle, an infinitude of such families, which
are, however, mutually inconsistent. This is addressed in this approach by the
so called \textquotedblleft single family rule\textquotedblright\ which
indicates one should never consider more than one family, at a time. One might
wonder, in fact, where does such rule come from?. In \cite{Weinberg-Col} it is
described as rather \textit{ad hoc}, but lets not focus on that issue here.
The issue we shall be concerned with, is the following: The need to single out
one particular family or realm providing the alternatives for the particular
history that becomes actual. The fact that one assigns probabilities within a
family, strongly suggests that the interpretation must be that one of the
histories in that family is actualized in our world. Otherwise, one must
wonder what these probabilities refer to (i.e. the probabilities assigned are
probabilities of what? (see however \cite{Hartle-Languaje}). Recall that, in
the context of the present problem we do not adopt the position that these are
probabilities of observing a certain value of a physical quantity when that
quantity is measured, because, as already discussed, we do not want to bring
concepts like measurement or observation into the discussion. In other words,
there is, in principle, no clear way to single out a specific family without
relying on an \textit{a-priori} given set of questions one is asking-- those
associated with the quantities whose spectral family one choses to construct
the realm - and this leads to serious interpretational difficulties
\cite{Okon}.

In the Mini-Mott experimental setup, we might guide ourselves, in practice, by
the questions the experimental set up is \textquotedblleft asking" (in fact,
this has a close analogy with the use of Bohr's rule in a given experiment or
series of experiments). However, in the absence of such guidance (i.e. without
\textit{a priori} considering that the experimental set up corresponds to
asking certain yes /no questions, as it seems to be required if one does
acknowledge the possibility of all superposition states of the apparatus
itself, and, in particular, taking the second basis for the discussion of the
Mini Mott example) one does not know how to select the family. Note that one
is not asking how to select a particular history within the family, but how to
select a particular family from within the collection of all possible
decoherent families.

In fact, it is hard to see, in describing the universe, what would dictate the
selection of the appropriate projector operators, and thus of the appropriate
family, (if we require a description which do not makes use of our own
existence and our own asking of certain questions, as part of the input).


This very issue makes its appearance in the cosmological context we are
concerned with. In fact, the problem can be seen clearly in the following
example: Consider the family of projector operators, where the chosen
projectors are not tied to the symmetry, as it is done in
\cite{HartleCosmologyNew}, and consider their results. Those might seem
satisfactory in connection to what one needs to understand, namely the shape
and a amplitude of the primordial spectrum of cosmological anisotropies and
inhomogeneities. However the point is that we could , alternatively, analyze
the situation by considering the following family: Construct the projector
operator into the space of homogeneous and isotropic states $P_{HI}$. This is
simply the projector into the intersection of the kernels of the generators of
translations and rotations. Next, we define $P_{non}\equiv I-P_{HI}$ the
orthogonal projector. It is clear that these are projector operators and
satisfy $P_{HI}+P_{non}=1$. Next, we take the initial state for the quantum
fluctuations (usually called the vacuum) $|\Phi_{0}\rangle$, and note that it
is homogeneous and isotropic.

Now take any set of values for time $\{t_{i}\}$ and consider the family
associated with that initial state and the two projector operators $P_{HI}$
and $P_{non}$ at all those times. This can easily be seen to define a family
of consistent histories, simply because the dynamics preserves the symmetries
of homogeneity and isotropy.

Thus, one might consider the following question, what is the probability that
(at a given time, characterized in the appropriate relational way), the
universe is isotropic and homogeneous?. This can be evaluated using the
formula (\ref{ProbCH}) starting with the vacuum state.

It is easy to see that any history containing the orthogonal projector at any
time $P_{non}$, will have zero probability, but the history containing only
the operators $P_{HI}$ will have probability one. This leads us to conclude
that, at any time, the universe is homogeneous and isotropic. It thus can have
no inhomogeneities or anisotropies at all.

Thus, in attempting to take this approach, we would have to face two problems.
One, the conclusion that our universe is and has always been homogenous and
isotropic, and, thus, we could not be here, but also the fact that the
approach has led us to two contradicting conclusions. This later one, and the
one obtained in, say, \cite{HartleCosmologyNew}. How could we chose trust one
of them and not the other, despite the fact that both are obtained by the very
same procedure\footnote{According to \cite{Hartle-Languaje} the posture is
that one should believe both, and use the appropriate one in connection with
the questions one is asking. This posture is not shared by other authors, for
instance \cite{HistContextuales}. Moreover it seems the implicit views
regarding the nature of science are very problematic in general (see for
instance \cite{CHCritics, Okon}).}?

In the case of Generalized contexts formalism, we have indicated that this
formalism can be considered as an alternative to the theory of consistent
histories, having the advantage that some problematic histories of the
consistent theory are eliminated. However, concerning the cosmological
problem, the formalism of generalized contexts has the same problems of the
theory of consistent histories: it does not give any rule for selecting a
privileged generalized context.

So far, we have argued that none of the widely used
interpretations\footnote{There exists many variants of the major themes we
have considered here, and they have not been described in detail because the
differences have no bearing on the issue at hand. Namely these variants fail
to address the issue we face, for exactly the same reasons as the major ones
they are \textbf{closely connected } with. However, we acknowledge that there
might exist some other proposal we are unaware of, and which fare better in
dealing with the problem we have been considering in this work.} of quantum
theory can offer a satisfactory account of the question of the emergence of
primordial inhomogeneities out of quantum uncertainties in the inflationary universe.

\section{Possible paths to addressing the problem.}

As we have seen, there seems to be no good option\footnote{With the possible
exception of the d' Broglie- Bohm approach, where the source of the primordial
asymmetries is found in the initial conditions.}, within the existing range of
interpretational approaches to quantum theory, capable of addressing the issue
at hand. Essentially, we need a scheme that manages to evade the following
theorem: Given an initial state possessing a certain symmetry, and evolving
autonomously and deterministically with a dynamics respecting said symmetry,
one can not end up with a state that fails to have the symmetry in question.
Here, the central issue is that, in the cosmological setting in question,
there seems to be no way to call upon anything external to disrupt the
autonomous evolution. The reader might be surprised to see the word
deterministic in connection with quantum theory, but the fact is that
Shr\"odingier's equation is fully deterministic, and the only place where
determinism is lost in the context of quantum theory, is at the point where
one addresses the connection with the measurements. In the cosmological
setting at hand, as we have explained, we simply can not rely on concepts tied
to measurements. Thus, we are driven to look for solutions in the class of
theories that seek to modify quantum theory by assuming a generic departure
from Shr\"dingier's deterministic evolution, (i.e. even in the absence of
measurements). These are known generically as dynamical collapse theories, and
have been proposed as means to address the general measurement problem in
quantum theory. The most widely known examples are \cite{Diosi, Penrose, GRW,
Pearle89, Bassi, CSL} and the best known advocate of such ideas has been R.
Penrose, joined recently by S. Weinberg\cite{Weinberg-Col}.

Thus, the path seems to require the extension of these dynamical collapse
theories to the inflationary cosmology regime, an extension that requires both
the application of the ideas to quantum field theory, rather than to non
relativistic quantum mechanics, as well as the incorporation of gravitation
into the picture. The approach followed in the first treatments of this
problem has been rather simplistic: Introduce a one time spontaneous and
random collapse per mode of the quantum field taking place during the
inflationary regime\cite{Us}. The main idea has been to consider the
predictions emerging from the proposals in order to find the particular
assumptions needed so as to obtain a broad agreement with the observational
data\cite{UsPhen} . There are now several works that have been based on a
particular proposal for the class of dynamics controlling that hypothetical
dynamical collapse. In particular, the proposal known as Continuous
Spontaneous Localization\cite{CSL} has been adapted, in various forms, to the
problem at hand leading to conclusions that depend on the particular scheme of
adaptation used\cite{CSL-Inflation}.

Moreover we note, in relation to the Mini Mott example discussed in section
III, that a common feature of these \textquotedblleft dynamical collapse
theories" is that they privilege position variables ( or other closely
connected objects tied to localization) over other variables. In fact, the
basic objective of these theories could be roughly characterized as modifying
quantum theory to prevent the existence (or the extended persistence) of
quantum superpositions representing macroscopic objects localized at
macroscopically different places. Thus, when applying those theories to the
Mini Mott example, one would find that the first basis for the description of
the states of the detectors is preferred over the second, symmetric basis,
simply because the different relations of the two with the position variables.
Thus those theories would lead to collapse in the first basis (or something
very close to that) and not in the symmetric basis. Therefore, these theories
could account for the breakdown in the symmetry. Such account could be then
characterized heuristically by saying that the symmetry in question was
incompatible with the localization which is a feature of the states that are
broadly stable under collapse. The mathematics of the theories, of course,
would reflect that heuristic characterization. The same can not be said of any
of the interpretational approaches to quantum theory except for the de-Brogile
Bohm proposal that shares with the collapse theories the privileged status
given to the position variables.

It is also noteworthy the fact discussed in \cite{Benefits}, that
\textquotedblleft dynamical collapse theories" seem to offer paths to resolve
long standing issues afflicting proposals for quantum theories of gravitation,
such as the \textquotedblleft black hole information paradox" associated with
their expected evaporation through Hawking radiation, and \textquotedblleft
the problem of time" in quantum gravity.

At this point, we should caution the reader that those approaches are still in
the development stage. In particular, before any of those can be considered as
a serious contender for fully resolving the problem, they would have to be
made into a close and self consistent modification of quantum theory in
general (i.e. they should cover, in a unified manner, the many particle
systems addressed by the GRW or CSL proposals, and those requiring field
theoretical and general relativistic treatment such as the cosmological
problem we have been considering) , and in connection with its applicability
to these filed theoretical and gravitational contexts, it would have to face
the difficulties connected with issues of covariance, as well as conservation
laws. It is nevertheless worth pointing out that there are promising
developments on these topics, such as \cite{Lorentz Inv} and \cite{Energy
Conservation}.

\section{Discussion}

We have reviewed the various issues related to the interpretation of quantum
mechanics, and, in particular, the \textquotedblleft measurement
problem\textquotedblright, using as a guide the process of generation of
structure from quantum fluctuations in inflationary cosmology. The discussion
of the conceptual issues was facilitated by considering, within the same
conditions associated with that cosmological problem, the paradigmatic problem
where symmetry serves as a focal point: The quantum decay of a nucleus in a
spherically symmetric initial state leading to the well known straight traces
in a bubble chamber, a problem studied by Sir.N. F. Mott in the early days of
quantum theory.

We have, in fact, focused our attention on a simplified version of Mott's
problem, which allows not only an explicit writing of the complete system's
(particle and detectors) hamiltonian and quantum mechanical states, but also
an explicit solution of the Shr\"odinger equation. This has made it possible
to investigate the problem in all detail. This has offered us a clear way to
exhibit the strengths and weaknesses of each one of the proposed
interpretations of quantum theory.

In Section IV, we have seen how each one of these interpretations fare in the
face of the cosmological problem we had described:

\begin{enumerate}
\item[a.] The Classical Apparatuses Interpretation does not solve the
cosmological problem because it needs the introduction of external (classical) detectors.

\item[b.] The Copenhagen Interpretation faces the same situation.

\item[c.] The Statistical Interpretation does not solve the cosmological
problem because, in this context, there are no measuring apparatuses, and no
measurements. These interpretations fail to deal with our problem, basically
from the start simply because in the cosmological context we need to account
for the emergence of the primordial inhomogeneities and anisotropies which are
the seeds of all cosmic structure, including galaxies, stars, planets, life,
humans (or other sapient beings), and instruments. Thus we have to do without
instruments at the stage where we want to understand the emergence of those
primordial features.

\item[d.] In Modal interpretations the symmetric initial state evolves with a
symmetric Hamiltonian, and then the symmetry will be present in the state at
all times. In addition:

\begin{enumerate}
\item[i.] The Atomic Modal Interpretation does not offer a choice of the
privileged base.

\item The Biorthogonal-decomposition modal interpretation requires for its
application a privileged basis sellected beforehand.

\item The Perspectival modal interpretation faces the same problems as those
appearing when attempting to rely on discussion based on decoherence.

\item The Modal-Hamiltonian interpretation can not be applied to systems with
a time-dependent Hamiltonian.
\end{enumerate}

\item[e.] The de-Brogile Bohm Interpretation.Using it we can not really argue
that it leads to a breakdown of the initial symmetry, either in the Mini-Mott
problem, or in the cosmological context where one wants to explain the
emergence to the seeds of cosmic structure, because, with n this approach the
symmetry was never there to start with. That is, even though the wave function
is symmetric, the initial values for the position variables are not symmetric.
However, we must acknowledge that although one does not have the right to
argue that this approach explains the emerge of asymmetry, it seems to be able
to account for the seeds of structure in the universe (provided that one
assumes that the preferential variable is something like the field amplitude
and that the initial condition corresponds, in some particular sense, to
something close to the equilibrium distribution \cite{CosmoBohm}. With such
additions, this approach seems to be the only competition to the one described
in the previous section.

\item[f.] In the case of Many Worlds Interpretation we saw that the schemes
posseses no elements allowing one to select the basis, or context in which the
ensemble must be described, or in which the splitting of the world takes
place, and that depending on the arbitrary choice that one makes, one could be
led to argue in favor or against the breakdown of the initial symmetry, and
thus in favor or against of the emergence of structure in our universe.

\item[g.] Something similar happens with the Interpretations based on
histories. One simply does not have anything like an unambiguous rule
indicating which kind of realm to consider, and depending on the choice, one
might end up assigning non-vanishing probabilities to non symmetric histories,
or an exactly vanishing probability for all except the symmetric ones.
\end{enumerate}

Thus we conclude that none of the existing interpretational frameworks for
quantum theory offers a satisfactory account leading to the desired breaking
of the initial symmetry in the problem at hand, and leading to what we think
are the appropriate characterization of the late time situations where the
symmetry is gone.

The analysis presented here indicates that something new is required, and we
have briefly scketched what we feel is a promising path in the search for a
clear characterization of that novel aspect of physics: dynamical collapse theories.

\section*{Acknowledgments}

\noindent This work was supported, in part, by CONACYT (M\'exico) Project
101712, a PAPPIT-UNAM (M\'exico) project IN107412 and sabbatical fellowships
from CONACYT and DGAPA-UNAM (M\'exico). D.S. wants to thank the IAFE at the
university of Buenos Aires for the hospitality during the sabbatical stay.
This work was partially supported by grants: of the Research Council of
Argentina (CONICET), by the Endowment for Science and by Technology of
Argentina (FONCYT), and by the University of Buenos Aires. We acknowledge very
useful discussions with B. Kay and Elias Okon.

\section*{APPENDIX}

In this appendix, we discuss some specific issues that arise in the attempt to
use decoherence related arguments in the context of the problem at hand.

The first issue is that connected to the implication of symmetry regarding the
choice of a preferential basis or so called pointer states.

The simplest example exhibiting this problem is provided by a standard EPR-R
setup: Consider the decay of a spin 0 particle into two spin 1/2 particles.
Take the direction of the decay as being the $x $ axis (the particles momenta
are $\vec P= \pm P \hat{x}$ with $\hat{x}$ the unit vector in the $\vec x $
direction)\footnote{We are ignoring here the issue of how this decay became
actualized into that particular direction, as the point here is to exemplify a
specific technical issue.}. Now, we characterize the two particle states, that
emerges after the decay in terms of the $\vec{z}$ polarization states of the
two Hilbert spaces of individual particles. As it is known, the conservation
of the angular momentum of the system indicates that the state must be:%

\begin{equation}
|\chi\rangle= \frac{1}{\sqrt2}( |+ \rangle^{(1)}_{z} |- \rangle^{(2)}_{z} + |+
\rangle^{(2)}_{z} |- \rangle^{(1)}_{z} )
\end{equation}

The state is clearly invariant under rotations around the $x$ axis (simply
because it is an eigen-state with zero angular momentum along that axis). The
density matrix for the system is thus $\rho= |\chi\rangle\langle\chi|$. Now
assume we decide we are not interested in one of the particles (call it $1$),
and thus we regard it as an environment for the system of interest (particle
$2$). The reduced density matrix is then:%

\begin{equation}
\rho^{(2)} = \rho_{Reduced}= Tr_{(1)} \rho= \frac{1}{ 2}( |+ \rangle^{(2)}_{z}
\langle+|^{(2)}_{z}+|- \rangle^{(2)}_{z} \langle-|^{(2)}_{z} )
\end{equation}

Now suppose we want to say that as the reduced density matrix is diagonal, we
have found the pointer basis and that somehow the particle must be considered
as having its spin along the $z$ axis defined to be either $+1/2 $ or $-1/2$.

The problem is that the symmetry of the state $|\chi\rangle$ regarding
rotations around the $x$ axis is reflected in the fact that we could have
written this density matrix also as
\begin{equation}
\rho^{(2)} = \frac{1}{ 2}( |+ \rangle^{(2)}_{y} \langle+|^{(2)}_{y}+|-
\rangle^{(2)}_{y} \langle-|^{(2)}_{y} )
\end{equation}
leading, this time, to the conclusion that the particle must be considered as
having its spin along the $y$ axis defined to be either $+1/2 $ or $-1/2$.

In fact, as the density matrix is proportional to the identity ( i.e.
$\rho^{(2)} = \frac{1}{ 2} I$) it would have the same form in any orthogonal basis.

One might be inclined to consider that this problem occurs only in very simple
situations, such as the one of the above example, and that, in general, we
will not encounter such difficulty. However that consideration is mistaken as
can be seen from the general result encapsulated in the following:

\subsection{ \textbf{Theorem:}}

\medskip Consider a quantum system made of a subsystem $S$ and an environment
$E$, with corresponding Hilbert spaces $H_{S}$ and $H_{E}$ so that the
complete system is described by states in the product Hilbert space
$H_{S}\otimes H_{E}$. Let $G$ be a symmetry group acting on the Hilbert space
of the full system in a way that does not mix the system and environment. That
is, the unitary representation $O$ of $G$ on $H_{S}\otimes H_{E}$ is such that
$\forall g \in G$, $\hat O(g) = \hat O^{S}(g)\otimes\hat O^{E}(g)$, where
$\hat O^{S}(g)$ and $\hat O^{E}(g)$ act on $H_{S}$ and $H_{E}$ respectively.

Let the system be characterized by a density matrix $\hat\rho$ which is
invariant under $G$. Then the reduced density matrix of the subsystem is a
multiple of the identity in each invariant subspace of $H_{S}$.

\subsection{Proof}

The reduced density matrix $\hat\rho_{S}= Tr_{E} ( \hat\rho)$.The trace over
the environment of any operator $\hat A$ in $H_{S}\otimes H_{E}$ is obtained
by taking any orthonormal basis $\lbrace|e_{j}\rangle\rbrace$ of $H_{E}$ and
evaluating $\Sigma_{j} \langle e_{j}| \hat A |e_{j}\rangle$. \medskip

Now, by assumption, we have $\hat\rho= {\hat O(g)}^{\dagger}\hat\rho\hat
O(g)$, $\forall g \in G$. Then, for all $g \in G$, we have $\hat\rho_{S} =
\Sigma_{j} \langle e_{j}| \hat\rho|e_{j}\rangle= \Sigma_{j} \langle e_{j}|
{\hat{O}^{S}(g)}^{\dagger}\otimes{ \hat{O}^{E}(g)}^{\dagger}\hat\rho\hat
O^{S}(g)\otimes\hat O^{E}(g) |e_{j}\rangle= \Sigma_{j} {{\hat O}^{S}%
(g)}^{\dagger}\langle e^{\prime}_{j}| \hat\rho|e^{\prime}_{j}\rangle{\hat
O}^{S}(g)$, where $|e^{\prime}_{j}\rangle\equiv O^{E}(g) |e_{j}\rangle$.
However, the fact that the operator ${\hat O}^{E}(g)$ is unitary implies that
the transformed states $\lbrace|e^{\prime}_{j}\rangle\rbrace$ form also an
orthonormal basis of $H_{E}$. \medskip

Thus we have $\hat\rho_{S} = {{\hat O}^{S}(g)}^{\dagger}( \Sigma_{j} \langle
e^{\prime}_{j}| \hat\rho|e^{\prime}_{j}\rangle) {\hat O}^{S}(g ) = {{\hat
O}^{S}(g)}^{\dagger}\hat\rho_{S} {\hat O}^{S}(g )$ or equivalently $\hat
\rho_{S} {{\hat O}^{S}(g)} = {\hat O}^{S}(g ) \hat\rho_{S}$. So we have found
that $[ \hat\rho_{S} , {\hat O}^{S}(g )]=0$, $\forall g \in G$, and thus by
Schur's lemma it follows that $\hat\rho_{S}$ must be a multiple of the
identity in each invariant subspace of $H_{S}$, QED. \medskip

In particular, this result indicates that, if we start with a pure state
invariant under the symmetry group, the reduced density matrix must be a
multiple of the identity in each invariant subspace of $H_{S}$. This is
exemplified by the well known case of a standard EPR setting, where a spinless
particle decays into two photons, and where one considers the photons' spin
degrees of freedom. The reduced density matrix describing one of the photons
is a multiple of the identity, and thus the decoherence that results from
tracing over the first photon's spin does not determine a preferential basis
for the characterization of the spin of the second photon. Decoherence then
fails under these conditions to provide a well defined preferential context
for the interpretation of the reduced density matrix, as representing the
various alternatives for the state of the subsystem after decoherence.


\begin{thebibliography}{99}                                                                                               %
\bibitem {QGTime}B. S. De Witt \textit{Phys. Rev.} \textbf{160}, 1113,(1967);
J.A. Wheeler in \textit{Battelle Reencontres 1987} eds. C. De Witt \& J.A
Wheeler (Benjamin, New York,1968); ``Canonical Quantum Gravity and the problem
of Time", C.J. Isham \textbf{GIFT Semminar}- 0157228 (1992) qr-qc/9210011.

\bibitem {Conceptual Problems  of QG}
see for instance J. Isham, (1992). gr-qc/9210011.

\bibitem {Problems Big Bang}``Inflationary universe: A possible solution to
the horizon and flatness problems"A. Guth \textit{Phys. Rev. D}\textbf{23},
347, (1981). For a more exhaustive discussion see for instance the relevant
chapter in ``The Early Universe", E.W. Kolb and M.S. Turner, Frontiers in
Physics Lecture Note Series (Addison Wesley Publishing Company 1990).



\bibitem {HZ}
``Fluctuations at the threshold of classical cosmology" E. R. Harrison,
\textit{Phys. Rev. D}, \textbf{1}, 2726, (1970); Y. B. Zel\'{}dovich
\textit{Mon. Not. Roy. Astron. Soc.} \textbf{160}, 1p (1972).

\bibitem {CMB exp}``Cosmological parameters From First results of Boomerang"
A. E. Lange \textit{et. al.} \textit{Phys. Rev. D}, \textbf{63}, 042001,
(2001); G. Hinshaw \textit{et. al.}, \textit{Astrophys. J. Supp.},
\textbf{148}, 135, (2003); ``Power Spectrum of Primordial Inhomogeneity
Determined from four Year COBE DMR SKY Maps", K. M. Gorski \textit{et. al.}
\textit{Astrophys. J.} \textbf{464}, L11, (1996); ``First Year Wilkinson
Microwave Anisotropy Probe (WMAP) Observations: Preliminary Results" C. L.
Bennett \textit{et. al.} \textit{Astrophys. J. Suppl.} \textbf{148}, 1,
(2003); ``First Year Wilkinson Microwave Anisotropy Probe (WMAP) Observations:
Foreground Emission", C. Bennett \textit{et. al.} \textit{Astrophys. J.
Suppl.} \textbf{148}, 97, (2003); G.~Hinshaw \textit{et al.} [WMAP
Collaboration], Nine-Year Wilkinson Microwave Anisotropy Probe (WMAP)
Observations: Cosmological Parameter Results, arXiv:1212.5226 [astro-ph.CO];
D.~Larson, J.~Dunkley, G.~Hinshaw, E.~Komatsu, M.~R.~Nolta, C.~L.~Bennett,
B.~Gold and M.~Halpern \textit{et al.}, Seven-Year Wilkinson Microwave
Anisotropy Probe (WMAP) Observations: Power Spectra and WMAP-Derived
Parameters, Astrophys.\ J.\ Suppl.\ \textbf{192}, 16 (2011) [arXiv:1001.4635
[astro-ph.CO]]; P.~A.~R.~Ade \textit{et al.} [Planck Collaboration], Planck
2013 results. XV. CMB power spectra and likelihood, arXiv:1303.5075 [astro-ph.CO].

\bibitem {Trashen}\textquotedblleft Semiclassical physics and quantum
fluctuations", W. Boucher \& J. Traschen, \textit{Physical Review D}
\textbf{37}, pp. 3522-3532, \ 1988.

\bibitem {Muckhanov}Page 348 of ``Physical Foundations of Cosmology", V.
Muckhanov ( Cambridge University Press, 2005)

\bibitem {Cosmologists}\textquotedblleft Decoherence in Quantum Cosmology",
J.J. Halliwell, \textit{Phys. Rev. D}, \textbf{39}, 2912,(1989);
\textquotedblleft Origin of Classical Structure From Inflation", C. Kiefer
\textit{Nucl.\ Phys.\ Proc.\ Suppl.\ } \textbf{88}, 255 (2000)
[arXiv:astro-ph/0006252]; \textquotedblleft Semiclassicality and decoherence
of Cosmological perturbations", D. Polarski and A.A.
Starobinsky,\textit{Class.\ Quant.\ Grav.\ } \textbf{13}, 377 (1996) [arXiv:
gr-qc/9504030]; \textquotedblleft\ Environment Induced Superselection In
Cosmology", W.H. Zurek, Environment Induced Superselection In Cosmology in
\textit{Moscow 1990, Proceedings, Quantum gravity} (QC178:S4:1990), p.
456-472. (see High Energy Physics Index 30 (1992) No. 624); \textquotedblleft
Gauge Invariant Cosmological Perturbations" R. Branderberger H. Feldman and V.
Mukhavov,\textit{Phys. Rep.} \textbf{215}, 203, (1992); \textquotedblleft
Decoherence Functional and Inhomogeneities in the Early Universe", R. Laflamme
and A. Matacz \textit{Int.\ J.\ Mod.\ Phys.\ D } \textbf{2}, 171 (1993)
[arXiv:gr-qc/9303036]; \textquotedblleft The self-induced approach to
decoherence in cosmology,\textquotedblright\ M. Castagnino and O. Lombardi,
\textit{Int. J. Theor. Phys.} \textbf{42}, 1281, (2003),
[arXiv:quant-ph/0211163]; \textquotedblleft Decoherence during inflation: The
generation of classical inhomogeneities,\textquotedblright\ F.~C.~Lombardo and
D.~Lopez Nacir, Phys.\ Rev.\ D \textbf{72}, 063506 (2005)
[arXiv:gr-qc/0506051]; \textquotedblleft Inflationary Cosmological
Perturbations of Quantum Mechanical Origin" J. Martin, \textit{Lect.\ Notes
Phys.\ } \textbf{669}, 199 (2005) [arXiv:hep-th/0406011]; \textquotedblleft%
\ Best Unbiased Estimates for Microwave background Anisotropies", L.P.
Grishchuk and J. Martin, \textit{Phys.\ Rev.\ D} \textbf{56}, 1924 (1997)
[arXiv:gr-qc/9702018]; \textquotedblleft Decoherence in Quantum Cosmology at
the onset of Inflation", A.O. Barvinsky, A.Y. Kamenshchik, C. Kiefer, and I.V.
Mishakov, \textit{Nucl.\ Phys.\ B} \textbf{551}, 374 (1999) [arXiv:gr-qc/9812043];

\bibitem {Padmanabhan}Section 10.4, page 364 of `` Structure Formation in the
Universe", T. Padmanabhan (Cambridge University Press, 1993).

\bibitem {Weinberg}Page 476 ``Cosmology", S. Weinberg ( Oxford University
Press 2008).

\bibitem {Mott}N. F. Mott \textquotedblleft The Wave Mechanics of $\alpha$-
Ray tracks", \textit{Proc. of the Royal Soc. of London}\textbf{126} No 800, pg
79, (1929).

\bibitem {decoherence}J. P. Paz \& W. H. Zurek, \textquotedblleft
Environment-induced decoherence and the transition from quantum to
classical\textquotedblright\textit{, }in D. Heiss (ed.), \textit{Lecture Notes
in Physics, Vol. 587} (Springer, 2002).

M. Schlosshauer, \textit{Decoherence and the Quantum-To-Classical Transition}
(Springer, 2007).

E. Joos \textit{et al}, \textit{Decoherence and the Appearance of a Classical
World in Quantum Theory} (Springer, 2003).

Castagnino M., Fortin S., (2011) International Journal of Theoretical Physics,
50 (7) , pp. 2259-2267.

Castagnino M., Fortin S., Lombardi O., (2010) Modern Physics Letters A, 25
(17) , pp. 1431-1439.

\bibitem {NO-SOL- MEASUREMENT}Butterfield, J. y Earman, J. (eds.) (2007),
\textit{Philosophy of Physics},\textit{\ Handbook of the Philosophy of
Science}, Amsterdam: North-Holland Elsevier.

\bibitem {Us}\textquotedblleft On the Quantum Mechanical Origin of the Seeds
of Cosmic Structure\textquotedblright\ A. Perez, H . Sahlmman, \& D. Sudarsky,
\textit{Classical and Quantum Gravity} \textbf{23} 2317 (2006); ``Towards a
formal description of the collapse approach to the inflationary origin of the
seeds of cosmic structure", Alberto Diez-Tejedor, \& Daniel Sudarsky,
\textit{JCAP}. \textbf{045}, 1207, (2012). e-Print: arXiv:1108.4928 [gr-qc].



\bibitem {Barbour}Julian B. Barbour, \textquotedblleft The timelessness of
quantum gravity: I. The evidence from the classical theory\textquotedblright,
Class. Quantum Grav. 11, 2853-2873, 1994.

Julian B. Barbour, \textquotedblleft The timelessness of quantum gravity: II.
The apperearance of Dynamics in statics configurations\textquotedblright,
Class. Quantum Grav. 11, 2853-2873, 1994.

John Earman, World Enough and Space-Time, Cambridge, MA: MIT Press, 1996.

\bibitem {QLogic}D. W. Cohen, \textit{An Introduction to Hilbert Space and
Quantum Logic}, Springer London, 2011.

Holik F., Massri C., Ciancaglini N., Int J Theo Phys (2012) 51: 1600-1620.

Holik F., Massri C., Plastino A., Zuberman L., Int J Theo Phys (2013) 52: 1836-1876.

\bibitem {FRASEN}B. C. van Fraassen, \textquotedblleft A formal approach to
the philosophy of science,\textquotedblright\ in Paradigms and Paradoxes: The
Philosophical Challenge of the Quantum Domain, R. Colodny (ed.), Pittsburgh:
University of Pittsburgh Press, pp. 303--366, 1972.

\bibitem {Holland}Peter R. Holland, \textit{The Quantum Theory of Motion: An
Account of the De Broglie-Bohm Causal Interpretation of Quantum Mechanics},
Cambridge University Press, 1995.

\bibitem {Bohm1952a}David Bohm, \textit{Physical Review} \textbf{85}, pp.
166-179, \ 1952.

\bibitem {Bohm1952b}David Bohm, \textit{Physical Review} \textbf{85}, pp.
180-193, \ 1952.

\bibitem {BohmDirac}David Bohm, \textit{Physical Review} \textbf{89}, pp.
458-466, \ 1953.

\bibitem {Zurek1982}W. H. Zurek, \textit{Physical Review D} \textbf{26},
1862-1880, 1982.

\bibitem {ScullySheaCullen}M. O. Scully, R. Shea, J. D. Mc Cullen. Phys. Rep.
43, 485-498 (1978).

\bibitem {Zurek1981}W. H. Zurek, Phys. Rev. D, 24, 1516-1525 (1981).

\bibitem {Valentini}A. Valentini,``Inflationary Cosmology as a Probe of
Primordial Quantum Mechanics" \textit{Physics Review D} \textbf{82}, 063513, (2010).

\bibitem {SEP-ManyWorlds}Vaidman, Lev, \textquotedblleft Many-Worlds
Interpretation of Quantum Mechanics\textquotedblright, \textit{The Stanford
Encyclopedia of Philosophy} (Fall 2008 Edition), Edward N. Zalta (ed.), URL =
$<$%
http://plato.stanford.edu/archives/fall2008/entries/qm-manyworlds/%
$>$%
.

\bibitem {SEP:Copenhagen}Faye, Jan, \textquotedblleft Copenhagen
Interpretation of Quantum Mechanics\textquotedblright, \textit{The Stanford
Encyclopedia of Philosophy} (Fall 2008 Edition), Edward N. Zalta (ed.), URL =
$<$%
http://plato.stanford.edu/archives/fall2008/entries/qm-copenhagen/%
$>$%
.

\bibitem {Ballentine}L. E. Ballentine, \textit{Quantum Mechanics}, New York:
Prentice Hall, 1990.

\bibitem {SEP-Decoherence}Bacciagaluppi, Guido, \textquotedblleft The Role of
Decoherence in Quantum Mechanics\textquotedblright, \textit{The Stanford
Encyclopedia of Philosophy} (Winter 2012 Edition), Edward N. Zalta (ed.), URL
=
$<$%
http://plato.stanford.edu/archives/win2012/entries/qm-decoherence/%
$>$%
.

\bibitem {Cohen}C. Cohen-Tannoudji, B. Diu and F. Lalo\"{e}, Quantum
Mechanics, John Wiley \& Sons, 1978.

\bibitem {SEP-Modal}Lombardi, Olimpia and Dieks, Dennis, \textquotedblleft
Modal Interpretations of Quantum Mechanics\textquotedblright, \textit{The
Stanford Encyclopedia of Philosophy} (Winter 2012 Edition), Edward N. Zalta
(ed.), URL =
$<$%
http://plato.stanford.edu/archives/win2012/entries/qm-modal/%
$>$%
.

\bibitem {ModalAtomic}Bacciagaluppi, G. and M. Dickson, \textit{Foundations of
Physics} \textbf{29}: 1165--1201, 1999.

\bibitem {ModalSmith}Kochen, S., 1985, \textquotedblleft A new interpretation
of quantum mechanics,\textquotedblright\ in Symposium on the Foundations of
Modern Physics 1985, P. Mittelstaedt and P. Lahti (eds.), Singapore: World
Scientific, pp. 151--169.

D. Dieks, \textit{Annalen der Physik} \textbf{7}: 174--190, 1988.

D. Dieks, \textit{Foundations of Physics} \textbf{38}: 1397--1423, 1989.

D. Dieks, \textit{Physics Letters A} \textbf{142}: 439--446, 1989.

\bibitem {ModalPerspectival}Bene, G. and D. Dieks, \textit{Foundations of
Physics} \textbf{32}: 645--671, 2002.

\bibitem {ModalHamiltoniana}O. Lombardi and M. Castagnino, \textit{Studies in
History and Philosophy of Modern Physics} \textbf{39}: 380--443, 2008.

\bibitem {HistConsistentes}R. B. Griffiths, \textit{Journal of Statistical
Physics} \textbf{36}: 219--272, 1984.

R. Omn\`{e}s, \textit{Reviews of Modern Physics} \textbf{53}: 893--932, 1988.

R. Omn\`{e}s, \textit{Reviews of Modern Physics} \textbf{57}: 357--382, 1989.

M. Gell-Mann and J. B. Hartle, \textquotedblleft Quantum Mechanics in the
Light of Quantum Cosmology\textquotedblright, in W. H. Zurek (ed.),
\textit{Complexity, Entropy, and the Physics of Information}, Reading, Mass.:
Addison-Wesley, pp. 425-458, 1990.

\bibitem {HistContextuales}R. Laura and L. Vanni, Found. Phys. \textbf{39}:
160-173, 2009.

\bibitem {mp}L. Vanni, R. Laura, Int. J. Theor. Phys. \textbf{52}, 2386-2394 (2013)

\bibitem {pr}M. Losada, L. Vanni, R. Laura, Phys. Rev. A \textbf{87}, 052128 (2013)

\bibitem {dp}M. Losada, R. Laura, Int. J. Theor. Phys. \textbf{52}, 1289-1299 (2013)

\bibitem {NoGoModal}G. Bacciagaluppi, International Journal of Theoretical
Physics 34: 1206-1215.

S. Kochen and E. P. Specker, Journal of Mathematics and Mechanics 17: 59-87, 1967.

R. Clifton, British Journal for the Philosophy of Science 47: 371-398, 1996.

P. E. Vermaas, Studies in History and Philosophy of Modern Physics 30:
403-431, 1999.

\bibitem {CosmoBohm}A. Valentini, `` Inflationary Cosmology as a Probe of
Primordial Quantum Mechanics", Phys. Rev. D \textbf{82}, 063513 (2010); N.
Pinto-Neto, G. Santos and W. Struyve, ``Quantum-to-classical transition of
primordial cosmological perturbations in de Broglie--Bohm quantum theory",
Phys. Rev. D \textbf{85}, 083506 (2012) [arXiv:1110.1339].

\bibitem {Hartle-Languaje}J.B. Hartle, ``Quantum Physics and Human Language",
{ J. Phys. A} \textbf{40}, 3101 (2007).

\bibitem {HartleCosmologyNew}``Quantum Cosmology Problems for the 21${}^{st}$
Century", J. B.Hartle, [e-Print: gr-qc/9701022]; ``Generalized Quantum
mechanics for Quantum Gravity", J. B. Hartle, [e-Print: gr-qc/0510126].

\bibitem {CHCritics}A. Kent, ``Consistent sets yield contrary inferences in
quantum theory", { Phys.Rev. Lett.} \textbf{87}, 15 (1997); F. Dowker and A.
Kent, ``On the consistent histories approach to quantum mechanics", { J.
Statist. Phys.} \textbf{82}, 1575 (1996) [arXiv:gr-qc/9412067]; A. Bassi and
G. C. Ghirardi, ``Can the decoherent histories description of reality be
considered satisfactory?", { Phys. Lett. A} \textbf{257}, 247 (1999)
[arXiv:gr-qc/9811050]; A. Bassi and G. C. Ghirardi, ``About the notion of
truth in the decoherent histories approach: A Reply to Griffiths", { Phys.
Lett. A} \textbf{265}, 153 (2000) [arXiv:quant-ph/9912065].

\bibitem {Okon}``On the Consistency of the Consistent Histories Approach to
Quantum Mechanics" E. Okon \& D. Sudarsky, \textit{Foundations of Physics}
\textbf{44}, 19-33 (2014), e-Print: arXiv 1301.2586.


\bibitem {Starobinski2}\textquotedblleft Quantum To Classical Transition of
Cosmological Perturbations for Non Vacuum Initial States", J. Lesggourges,
David Polarski and Alexei A. Starobinsky [arXiv: gr-qc/961101904030] .

\bibitem {Diosi}L. Diosi,``Gravitation and quantum mechanical localization of
macro-objects", Phys. Lett. A \textbf{105} , 199-202 (1984); L. Diosi,``A
universal master equation for the gravitational violation of quantum
mechanics", Phys. Lett. A \textbf{120}, 377 (1987); L. Diosi, ``Models for
universal reduction of macroscopic quantum fluctuations", Phys. Lett. A
\textbf{40}, 1165 (1989).

\bibitem {Penrose}R. Penrose, ``The Emperor's New Mind", Oxford University
Press, U.K. (1989) 480 p; R. Penrose, ``On Gravity's Role in Quantum State
Reduction", Gen. Rel. Grav. \textbf{28}, 581 (1996).

\bibitem {GRW}G.C. Ghirardi, A. Rimini, and T. Weber, ``A Unified Dynamics For
Micro And Macro Systems", Phys. Rev. D \textbf{34}, 470 (1986).

\bibitem {Pearle89}P.M. Pearle, ``Combining stochastic dynamical state-vector
reduction with spontaneous localization", Phys. Rev. A \textbf{39}, 2277 (1989).

\bibitem {Bassi}A.~Bassi and G.C.~Ghirardi, ``Dynamical reduction models'',
Phys.\ Rept.\ \textbf{379}, 257 (2003) [arXiv:quant-ph/0302164].

\bibitem {CSL}P. Pearle, ``Reduction of the State Vector by a Nonlinear
Schrodinger Equation", Phys. Rev. D \textbf{13}, 857 (1976); P. Pearle,
``Toward Explaining Why Events Occur", Int.J.Theor.Phys. \textbf{18}, 489
(1979); P. Pearle, ``Experimental tests of dynamical state-vector reduction",
Phys. Rev. D \textbf{29}, 235 (1984); P. Pearle, ``Combining Stochastic
Dynamical State Vector Reduction With Spontaneous Localization", Phys. Rev. A
\textbf{39}, 2277 (1989).

\bibitem {Weinberg-Col}S. Weinberg, ``Collapse of the State Vector",
UTTG-18-11, (2011) [arXiv:1109.6462].

\bibitem {UsPhen}``Phenomenological Analysis of Quantum Collapse as Source of
the Seeds of Cosmic Structure", A. de Unanue \& Daniel Sudarsky,
\textit{Physics Review D} \textbf{78}, pg.043510 (2008). arXiv:0801.4702 [gr-qc];

``The slow roll condition and the amplitude of the primordial spectrum of
cosmic fluctuations: Contrasts and similarities of standard account and the
``collapse scheme"". G. Le\'on Garc\'{\i}a \& D. Sudarsky, \textit{Classical
and Quantum Gravity} \textbf{27}, pg. 225017 (2010);

``Multiple quantum collapse of the inflaton field and its implications on the
birth of cosmic structure", G. Le\'on Garc\'{\i}a, A. De Unanue, \&D.
Sudarsky, \textit{Classical and Quantum Gravity}, \textbf{28}, 155010 (2011);
arXiv:1012.2419 [gr-qc];

``Novel possibility of observable non-Gaussianities in the inflationary
spectrum of primordial inhomogeneities", G. Le\'on Garc\'{\i}a, \& D.
Sudarsky, \textit{Sigma} \textbf{8}, 024, (2012);

``The collapse of the wave function in the joint metric-matter quantization
for inflation", A. Diez-Tejedor, G. Le\'on Garc\'{\i}a, \& D. Sudarsky,
\textit{Gen. Rel. \& Grav.} , \textbf{44}, 2965, (2012). e-Print:
arXiv:1106.1176 [gr-qc];

``Cosmological constraints on nonstandard inflationary quantum collapse
models" S. J. Landau, C. G. Scoccola, \& D. Sudarsky, \textit{Physics Review
D} \textbf{85}, 123001, (2012). arXiv:1112.1830 [astro-ph.CO].

\bibitem {CSL-Inflation}J. Martin, V. Vennin and P. Peter, ``Cosmological
Inflation and the Quantum Measurement Problem", (2012) [arXiv:1207.2086];

``CSL Quantum Origin of the Primordial Fluctuation", P. Ca\~nate, P. Pearl, \&
D. Sudarsky, \textit{Physics Review D}, \textbf{87}, 104024 (2013); e-Print: arXiv:1211.3463[gr-qc]



``Quantum to Classical Transition of Inflationary Perturbations - Continuous
Spontaneous Localization as a Possible Mechanism" S. Das, K. Lochan, S. Sahu,
\&T.P. Singh e-Print: arXiv:1304.5094 [astro-ph.CO].



\bibitem {Benefits}``Benefits of Objective Collapse Models for Cosmology and
Quantum Gravity" E. Okon \& D. Sudarsky, \textit{Foundations of Physics}
\textbf{44} 114-143, (2014)
arXiv:1309.1730v1 [gr-qc] .


\bibitem {Lorentz Inv}W. C. Myrvold, ``On peaceful coexistence: is the
collapse postulate incompatible with relativity?", Studies in History and
Philosophy of Modern Physics \textbf{33}, 435 (2002); R. Tumulka , ``On
spontaneous wave function collapse and quantum field theory", Proc. Roy. Soc.
Lond. \textbf{A 462}, 1897 (2006) [arXiv:quant-ph/0508230].

\bibitem {Energy Conservation}D. J. Bedingham, \textquotedblleft Relativistic
state reduction dynamics", Found. Phys. \textbf{41}, 686 (2011) [arXiv:1003.2774].

\bibitem {looming}Lombardi O., Fortin, S., Castagnino M.: The problem of
identifying the system and the environment in the phenomenon of decoherence,
in H. W. de Regt, S. Hartmann and S. Okasha (eds.), European Philosophy of
Science Association (EPSA). Philosophical Issues in the Sciences Volume 3,
Berlin: Springer, pp. 161-174 (2012).

\bibitem {looming2}Castagnino M., Fortin S. and Lombardi O., \textquotedblleft
Suppression of decoherence in a generalization of the spin-bath
model,\textquotedblright\ Journal of Physics A: Mathematical and Theoretical,
\textbf{43}: \# 065304 (2010).

\bibitem {MHIfields}Ardenghi J. S., Castagnino M. and Lombardi O.,
International Journal of Theoretical Physics \textbf{50},pp. 774-791 (2011).

\bibitem {Everett1957}Everett H., Reviews of Modern Physics \textbf{29},
454--462 (1957).
\end{thebibliography}
\end{document}